\documentclass[11pt,a4paper]{article}
\pdfoutput=1
\usepackage{jcappub}
\usepackage{ifthen}
\usepackage{epsfig}
\usepackage{booktabs} 
\usepackage{natbib}

\newcommand{\beqra}{\begin{eqnarray}}
\newcommand{\eeqra}{\end{eqnarray}}
\newcommand{\beq}{\begin{equation}}
\newcommand{\eeq}{\end{equation}}

\title{Prospects for direct detection of dark matter in an effective theory approach}
\author[a]{Riccardo Catena}
\affiliation[a]{Institut f\"ur Theoretische Physik, Friedrich-Hund-Platz 1, 37077 G\"ottingen, Germany}
\emailAdd{riccardo.catena@theorie.physik.uni-goettingen.de}

\abstract{We perform the first comprehensive analysis of the prospects for direct detection of dark matter with future ton-scale detectors in the general 11-dimensional effective theory of isoscalar dark matter-nucleon interactions mediated by a heavy spin-1 or spin-0 particle. The theory includes 8 momentum and velocity dependent dark matter-nucleon interaction operators, besides the familiar spin-independent and spin-dependent operators. From a variegated sample of 27 benchmark points selected in the parameter space of the theory, we simulate independent sets of synthetic data for ton-scale Germanium and Xenon detectors. From the synthetic data, we then extract the marginal posterior probability density functions and the profile likelihoods of the model parameters. The associated Bayesian credible regions and frequentist confidence intervals allow us to assess the prospects for direct detection of dark matter at the 27 benchmark points. First, we analyze the data assuming the knowledge of the correct dark matter nucleon-interaction type, as it is commonly done for the familiar spin-independent and spin-dependent interactions. Then, we analyze the simulations extracting the dark matter-nucleon interaction type from the data directly, in contrast to standard analyses. This second approach requires an extensive exploration of the full 11-dimensional parameter space of the dark matter-nucleon effective theory. Interestingly, we identify 5 scenarios where the dark matter mass and the dark matter-nucleon interaction type can be reconstructed from the data simultaneously. We stress the importance of extracting the dark matter nucleon-interaction type from the data directly, discussing the main challenges found addressing this complex 11-dimensional problem.}

\keywords{dark matter theory, dark matter experiments} 

\begin{document}
\maketitle

\section{Introduction}
Detecting the particles forming the Milky Way dark matter halo is one of the priority of astroparticle physics~\cite{appec}. The direct detection technique is currently playing an important role in this context~\cite{Strigari:2013iaa}. Its goal is to measure the energy deposited in an underground detector by dark matter particles of the galactic halo scattering off the detector nuclei~\cite{Goodman:1984dc}. Cryogenic bolometers and liquid noble gas scintillators are two important examples of detectors used in this field of research~\cite{Cerdeno:2014uga}. Current direct detection experiments have reached the sensitivity to probe a broad spectrum of possible dark matter-nucleon interactions, including those depending on the dark matter-nucleus relative velocity and the momentum transfer~\cite{Catena:2014uqa}. LUX, SuperCDMS and CDMSlite are currently setting the most stringent bounds on the velocity and momentum independent dark matter coupling to the Xenon and Germanium nuclear charge density operators~\cite{Akerib:2013tjd,Agnese:2014aze,Agnese:2013jaa}. 

The next generation of direct detection experiments will exploit ton-scale targets operating in a low background environment~\cite{Baudis:2012ig}. There are great expectations on the discovery potentials of these new experimental devices, since their large exposure will allow to probe the vast majority of the particle models for weakly interacting dark matter~\cite{Akrami:2010dn,Pato:2011de,Strege:2012kv,Catena:2013pka}. The prospects for direct detection of dark matter with ton-scale detectors have been analyzed using complementary approaches. Interesting investigations performed in this context include applications of the extended Likelihood approach \cite{Green:2008rd}, studies of the interplay of Bayesian and frequentists statistics~\cite{Trotta:2006ew,Akrami:2010cz}, analyses of the complementarity of different detection strategies~\cite{Bergstrom:2010gh} and target materials~\cite{Pato:2010zk,Arina:2013jma,Peter:2013aha}, attempts to reconstruct the local dark matter velocity distribution~\cite{Gondolo:2002np,Drees:2007hr,Peter:2011eu,Peter:2013aha}, and an exploration of the Sun's gravitational focusing effect~\cite{Bozorgnia:2014dqa}.

The prospects for direct detection of dark matter depend not only on the strength of the dark matter-nucleon interaction, but also on the momentum and velocity dependence of the scattering amplitude in the non-relativistic limit. The majority of the forecasts made regarding the prospects for direct detection of dark matter with ton-scale detectors assume that dark matter couples to the velocity and momentum independent nuclear charge density and spin current density operators only. The former interaction operator generates the so-called ``spin-independent'' interaction, the latter the familiar ``spin-dependent'' interaction. A broader set of momentum and velocity dependent interaction operators is however allowed by Galilean invariance, and energy and momentum conservation~\cite{Fitzpatrick:2012ix}. The prospects for detecting a dark matter signal produced by 4  momentum-dependent interaction operators and 5 linear combinations of non-relativistic operators have been recently studied in Refs.~\cite{Peter:2013aha} and~\cite{Gresham:2014vja}, respectively. An analysis extended to all momentum and velocity dependent interaction operators arising from the exchange of a heavy spin-0 or spin-1 particle, which explores the full multi-dimensional parameter space of the dark matter-nucleon interaction theory, is however still missing.

In this paper we perform the first comprehensive analysis of the prospects for direct detection of dark matter with future ton-scale detectors in the general 11-dimensional effective theory of isoscalar dark matter-nucleon interactions mediated by a heavy spin-one or spin-zero particle. Ref.~\cite{Fitzpatrick:2012ix} gives a systematic and complete formulation of this theory, extending the previous analysis of Ref.~\cite{Fan:2010gt}.  Within this theoretical framework, Refs.~\cite{Fitzpatrick:2012ib,Liang:2013dsa,Gresham:2014vja,Panci:2014gga} analyze the data of current direct detection experiments separately. Ref.~\cite{Catena:2014uqa} compares the full 11-dimensional parameter space of the theory to current observations in a global statistical analysis of several direct detection experiments, including the recent LUX, SuperCDMS and CDMSlite results. {\sffamily Mathematica}  packages to evaluate the relevant nuclear form factors~\cite{Anand:2013yka} and compute direct detection exclusion limits~\cite{DelNobile:2013sia} are publicly available. 

We draw our conclusions from a suite of synthetic data, that we generate from 27 benchmark points selected in the 11-dimensional parameter space of the model. From our synthetic data, we extract the profile likelihoods and the marginal posterior probability density functions of the model parameters. Using state-of-the-art Bayesian and frequentist statistical methods, we identify the most promising scenarios and outline the main challenges emerging from this study.

The paper is organized as follows. In Sec.~\ref{sec:theory} we introduce the non-relativistic effective theory of the dark matter-nucleon interaction. We also define the benchmark points in the parameter space of the theory, from which in Sec.~\ref{sec:data} we simulate independent samples of synthetic data. Sec.~\ref{sec:statistics} is devoted to the statistical methods used in the analyses of the synthetic data. The prospects for detecting dark matter with ton-scale detectors in the effective theory of the dark matter-nucleon interaction are presented in Sec.~\ref{sec:results}. Details on the dark matter response functions used in the calculations are provided in the Appendix. 

\section{The effective theory of the dark matter-nucleon interaction}
\label{sec:theory}
In this section we define the effective theory of the dark matter-nucleon interaction studied in the paper. A more complete introduction to the subject can be found in Refs.~\cite{Fan:2010gt,Fitzpatrick:2012ib,Fitzpatrick:2012ix,Anand:2013yka}.  

\subsection{Definitions}
The effective theory of the dark matter-nucleon interaction is a non-relativistic field theory where the interaction operators are restricted by Galilean invariance, energy and momentum conservation, and hermiticity~\cite{Fitzpatrick:2012ib}. In this framework, five non-relativistic Galilean invariant operators generate the algebra of $\chi$-nucleon effective interaction operators, where $\chi$ denotes the dark matter particle. The five operators are: the identity $1_\chi 1_N$, the momentum transfer $\vec{q}$, the $\chi$-nucleon transverse relative velocity operator $\vec{v}^{\perp}_{\chi N}$, and the dark matter and nucleon spin operators $\vec{S}_\chi 1_N$ and $1_\chi \vec{S}_{N}$, respectively. Any dark matter-nucleon interaction operator can be expressed as a combination of the five generating operators. In this paper, we restrict ourselves to dark matter-nucleon interactions arising from the exchange of a heavy spin-0 or spin-1 particle, and hence to the 10 operators listed in Tab.~1.\footnote{The additional operators $\mathcal{O}_{16}=-\mathcal{O}_{10}\mathcal{O}_{5}$,   $\mathcal{O}_{13}=\mathcal{O}_{10}\mathcal{O}_{8}$,  $\mathcal{O}_{15}=-\mathcal{O}_{11}\mathcal{O}_{3}$ and $\mathcal{O}_{14}=\mathcal{O}_{11}\mathcal{O}_{7}$ are difficult to generate in explicit particle models, whereas the remaining operator, $\mathcal{O}_2=(v^\perp_{\chi N})^2$, cannot be a leading-order operator in effective theories.}

\begin{table}[t]
    \centering
    \begin{tabular}{ll}
    \toprule
        $\mathcal{O}_1 = 1_{\chi} 1_{N}$ \hspace{10em} &         $\mathcal{O}_7 = \vec{S}_{N}\cdot \vec{v}^{\perp}_{\chi N}$ \\
        $\mathcal{O}_3 = -i\vec{S}_N\cdot\left(\frac{\vec{q}}{m_N}\times\vec{v}^{\perp}_{\chi N}\right)$ &         $\mathcal{O}_8 = \vec{S}_{\chi}\cdot \vec{v}^{\perp}_{\chi N}$ \\
        $\mathcal{O}_4 = \vec{S}_{\chi}\cdot \vec{S}_{N}$ &         $\mathcal{O}_9 = -i\vec{S}_\chi\cdot\left(\vec{S}_N\times\frac{\vec{q}}{m_N}\right)$ \\                                                                             
        $\mathcal{O}_5 = -i\vec{S}_\chi\cdot\left(\frac{\vec{q}}{m_N}\times\vec{v}^{\perp}_{\chi N}\right)$ &         $\mathcal{O}_{10} = -i\vec{S}_N\cdot\frac{\vec{q}}{m_N}$ \\                                                                                                        
        $\mathcal{O}_6 = \left(\vec{S}_\chi\cdot\frac{\vec{q}}{m_N}\right) \left(\vec{S}_N\cdot\frac{\vec{q}}{m_N}\right)$ &        $\mathcal{O}_{11} = -i\vec{S}_\chi\cdot\frac{\vec{q}}{m_N}$ \\                                                                                                      
    \bottomrule
    \end{tabular}
    \caption{List of the 10 non-relativistic operators defining the effective theory of the dark matter-nucleon interaction studied in this paper. The operators $\mathcal{O}_i$ are the same as in Ref.~\cite{Anand:2013yka}.}
    \label{operators}
\end{table}

The most general Lagrangian describing the dark matter-nucleon interaction is given by the linear combination 
\begin{equation}
\mathcal{L}_{\rm int} = \sum_{N={\rm n}, {\rm p}}\sum_{i} c_i^{N} \mathcal{O}_i \chi^+ \chi^- N^+ N^- \,,
\label{eq:L}
\end{equation}
where $\chi^+$ ($\chi^-$) and $N^+$ ($N^-$) are the positive (negative) frequency parts of the non-relativistic dark matter and nucleon fields, respectively. In Eq.~(\ref{eq:L}), $c^{\rm p}_i$ and $c^{\rm n}_i$ are the coupling constants for protons and neutrons. They are related to the isoscalar and isovector coupling constants $c^{\tau}_i$ ($\tau=0,1$) by the relations $c^{\rm p}_i = (c^{0}_i+c^{1}_i)/2$ and $c^{\rm n}_i = (c^{0}_i-c^{1}_i)/2$. Following Ref.~\cite{Anand:2013yka}, we introduce $c^\tau_i$ constants with dimension (mass)$^{-2}$. We collectively denote them by $\mathbf{c}$. In this paper we restrict our analysis to isoscalar interactions, i.e., we set $c^{1}_i=0$.  

The differential cross section for dark matter scattering on a target nucleus of mass $m_T$ is given by
\begin{equation}
\frac{d\sigma}{dE_{R}} = \frac{m_{T}}{2\pi v^2} \Bigg[ \frac{1}{2j_\chi+1}\frac{1}{2j_N+1} \sum_{\rm spins} |\mathcal{M}_{NR}|^2 \Bigg]
\label{dsigmadER}
\end{equation}
where $j_\chi$ and $j_N$ are, respectively, the dark matter and nucleus spins, while $\mathcal{M}_{NR}$ represents the non-relativistic scattering amplitude. We denote by $P_{\rm tot}$ the average of $|\mathcal{M}_{NR}|^2$ over initial spins, summed over final spins. $P_{\rm tot}$ is proportional to the total transition probability and it can be expressed as a combination of nuclear and dark matter response functions, namely 
\allowdisplaybreaks
\begin{eqnarray}
P_{\rm tot}({v}^2,{q}^2)&\equiv&{1 \over 2j_\chi + 1} {1 \over 2j_N + 1} \sum_{\rm spins} |\mathcal{M}_{NR}|^2 \nonumber \\  &=& {4 \pi \over 2j_N + 1} 
\sum_{ \tau=0,1} \sum_{\tau^\prime = 0,1} \Bigg\{ \Bigg[ R_{M}^{\tau \tau^\prime}({v}^{\perp 2}_{\chi T}, {{q}^{2} \over m_N^2})~W_{M}^{\tau \tau^\prime}(y)   \nonumber\\
&+& R_{\Sigma^{\prime \prime}}^{\tau \tau^\prime}({v}^{\perp 2}_{\chi T}, {{q}^{2} \over m_N^2})   ~W_{\Sigma^{\prime \prime}}^{\tau \tau^\prime}(y) 
+   R_{\Sigma^\prime}^{\tau \tau^\prime}({v}^{\perp 2}_{\chi T}, {{q}^{2} \over m_N^2}) ~ W_{\Sigma^\prime}^{\tau \tau^\prime}(y) \Bigg]  \nonumber\\  
&+& {{q}^{2} \over m_N^2} ~\Bigg[R_{\Phi^{\prime \prime}}^{\tau \tau^\prime}({v}^{\perp 2}_{\chi T}, {{q}^{2} \over m_N^2}) ~ W_{\Phi^{\prime \prime}}^{\tau \tau^\prime}(y)  +  R_{ \Phi^{\prime \prime}M}^{\tau \tau^\prime}({v}^{\perp 2}_{\chi T}, {{q}^{2} \over m_N^2})  ~W_{ \Phi^{\prime \prime}M}^{\tau \tau^\prime}(y) \nonumber\\
&+&   R_{\tilde{\Phi}^\prime}^{\tau \tau^\prime}({v}^{\perp 2}_{\chi T}, {{q}^{2} \over m_N^2}) ~W_{\tilde{\Phi}^\prime}^{\tau \tau^\prime}(y) 
+   R_{\Delta}^{\tau \tau^\prime}({v}^{\perp 2}_{\chi T}, {{q}^{2} \over m_N^2}) ~ W_{\Delta}^{\tau \tau^\prime}(y) \nonumber\\
 &+&  R_{\Delta \Sigma^\prime}^{\tau \tau^\prime}({v}^{\perp 2}_{\chi T}, {{q}^{2} \over m_N^2})  ~W_{\Delta \Sigma^\prime}^{\tau \tau^\prime}(y)   \Bigg]  \Bigg\}  \,.
\label{Ptot}
\end{eqnarray}
In Eq.~(\ref{Ptot}), 
\begin{align}
v^{\perp 2}_{\chi T} = v^2 - \frac{q^2}{4\mu_T^2} \,
\end{align}
where $v$ and $\mu_T$ are the dark matter-nucleus relative velocity and reduced mass, respectively.
The dark matter response functions $R_{M}^{\tau \tau^\prime}$, $R_{\Sigma^{\prime \prime}}^{\tau \tau^\prime}$, $R_{\Sigma^\prime}^{\tau \tau^\prime}$, $R_{\Phi^{\prime \prime}}^{\tau \tau^\prime}$, $R_{\Phi^{\prime\prime}M}^{\tau \tau^\prime}$, $R_{\tilde{\Phi}^\prime}^{\tau \tau^\prime}$, $R_{\Delta}^{\tau \tau^\prime}$ and $R_{\Delta \Sigma^\prime}^{\tau \tau^\prime}$ are listed in the Appendix. The nuclear response functions  $W_{M}^{\tau \tau^\prime}$, $W_{\Sigma^{\prime \prime}}^{\tau \tau^\prime}$, $W_{\Sigma^\prime}^{\tau \tau^\prime}$, $W_{\Phi^{\prime \prime}}^{\tau \tau^\prime}$, $W_{\Phi^{\prime\prime}M}^{\tau \tau^\prime}$, $W_{\tilde{\Phi}^\prime}^{\tau \tau^\prime}$, $W_{\Delta}^{\tau \tau^\prime}$ and $W_{\Delta \Sigma^\prime}^{\tau \tau^\prime}$ are defied in Eq.~(41) of Ref.~\cite{Anand:2013yka}. We evaluate the nuclear response functions using our {\sffamily FORTRAN} version of the {\sffamily Mathematica} package introduced in Ref.~\cite{Anand:2013yka}. In Eq.~(\ref{Ptot}), $y=(q b/2)^2$, where $b$ is the oscillator parameter in the independent-particle harmonic oscillator model~\cite{Anand:2013yka}.

The differential rate of scattering events per unit time and per unit detector mass is given by
\begin{equation}
\frac{{\rm d}\mathcal{R}}{{\rm d}E_{R}} =  \sum_{T}\frac{{\rm d}\mathcal{R}_{T}}{{\rm d}E_{R}} \equiv  \sum_{T} \xi_T \frac{\rho_{\chi}}{2\pi m_\chi}  \left\langle  \frac{1}{v} P_{\rm tot}(v^2,q^2)  \right\rangle
\label{rate_theory}
\end{equation}
where $m_\chi$ is the dark matter mass, $\rho_\chi$ is the local dark matter density and $\xi_T$ is the mass fraction of the nucleus $T$ in the target material. In Eq.~(\ref{rate_theory}), the angle brackets denote the average
\begin{equation}
\left\langle \frac{1}{v} P_{\rm tot}(v^2,q^2)  \right\rangle =  \int\limits_{v>v_{\rm min}(q)} \,  \frac{f(\vec{v} + \vec{v}_e(t))}{v} \, P_{\rm tot}(v^2,q^2) \, d^3v  , 
\label{eq:uDF}
\end{equation} 
where $f$ is the local dark matter velocity distribution in the galactic rest frame boosted to the detector frame. In Eq.~(\ref{eq:uDF}), $v_{\rm min}(q)=q/2\mu_T$ is the minimum velocity that a dark matter particle must have in order to transfer a momentum $q$ to the target nucleus, and $\vec{v}_e(t)$ is the time-dependent Earth velocity in the galactic rest frame. In this paper we consider the anisotropic velocity distribution proposed in Ref.~\cite{Bozorgnia:2013pua}, with astrophysical parameters set at their mean values, i.e. blue line in the left panel of Fig.~6 in Ref.~\cite{Bozorgnia:2013pua}. See also Refs.~\cite{Catena:2009mf,Catena:2011kv} for an introduction to this galactic model.

\begin{figure}[t]
\begin{center}
\includegraphics[width=\textwidth]{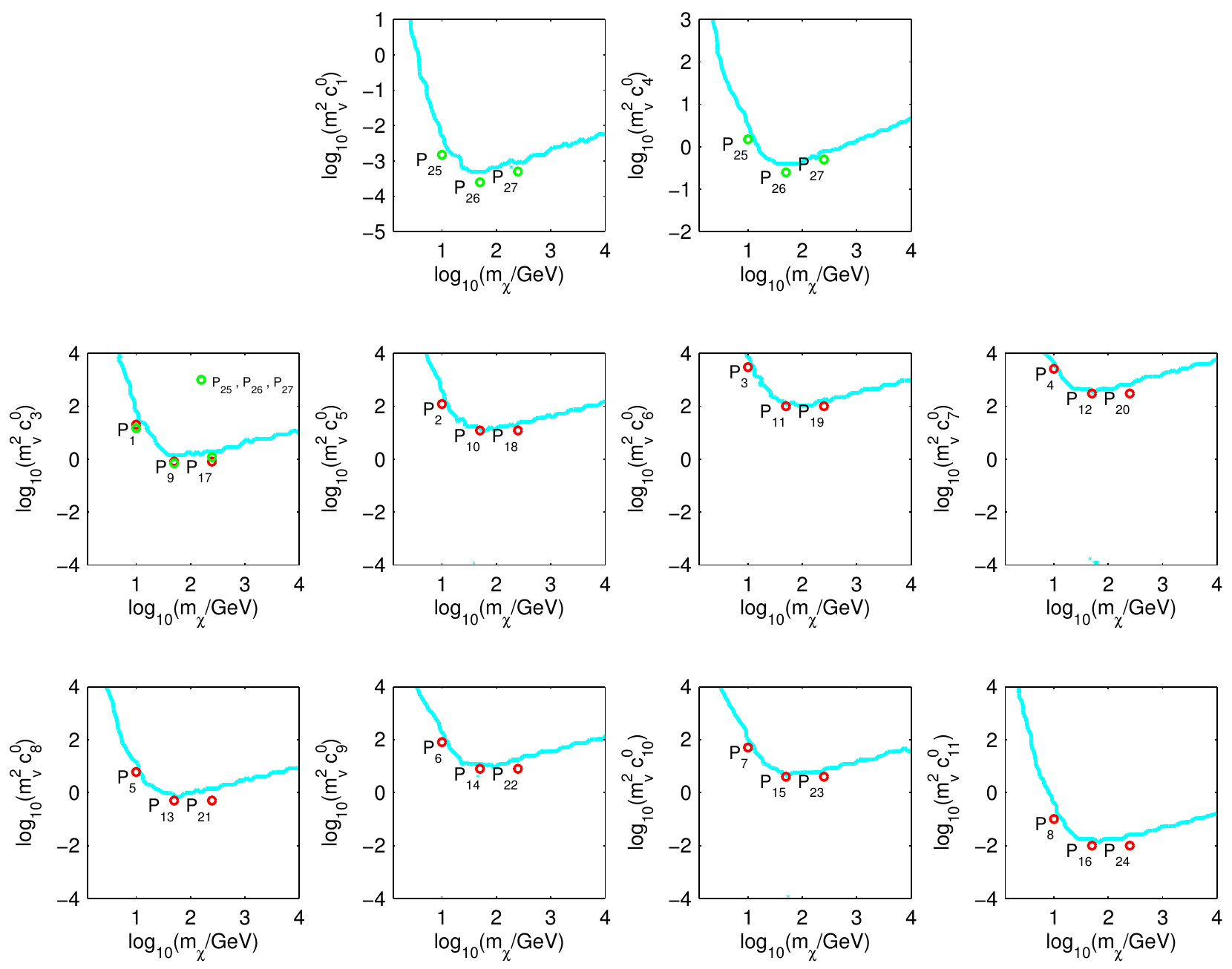}
\end{center}
\caption{Benchmark points selected in the 11-dimensional parameter space of the dark matter-nucleon effective theory studied in this paper. The 27 benchmark points are represented in the 10 planes $m_\chi$-$c_{i}^0$, with $i=1,3,\dots,11$. At the benchmark points $P_{25}$, $P_{26}$ and $P_{27}$, dark matter interacts with the nucleons through a linear combination of the operators $\mathcal{O}_1$, $\mathcal{O}_3$ and $\mathcal{O}_4$. At the other benchmark points, a single operator is responsible for the dark matter-nucleon interaction. The properties of the benchmark points are listed in Tab.~\ref{tab:benchmarks}. The cyan contours in the figure represent exclusion limits at the 95\% confidence level on the coupling constants of the dark matter-nucleon effective theory. These limits have been derived in Ref.~\cite{Catena:2014uqa} through a global analysis of current direct detection experiments.}
\label{fig:benchmarks}
\end{figure}

\subsection{Benchmark points $P_1$ -- $P_{27}$}
\label{sec:benchmarks}
The effective theory of isoscalar dark matter-nucleon interactions mediated by heavy spin-one or spin-zero particles depends on the 10 constants $c_i^0$, $i=1,3,\dots,11$, besides the dark matter mass. In order to assess the prospects for direct detection of dark matter in this theoretical framework, we have selected 27 benchmark points in the parameters space of the theory. Their properties are summarized in Tab.~\ref{tab:benchmarks}, and illustrated in the 10 planes $m_\chi$-$c_i^0$, $i=1,3\dots,11$, in Fig.~\ref{fig:benchmarks}. We have selected 8 light benchmark points, $P_{1}$ -- $P_{8}$ with $m_\chi=10$~GeV; 8 relatively heavy benchmark points, $P_{17}$ -- $P_{24}$ with $m_\chi=250$~GeV; and 8 benchmark points with $m_\chi=50$~GeV, namely $P_{9}$ -- $P_{16}$. As shown in Tab.~\ref{tab:benchmarks}, the benchmark points $P_1$ -- $P_{24}$ allow us to analyze the prospects for detecting a signal produced by the interaction operators $\mathcal{O}_3$, $\mathcal{O}_5$,\dots, $\mathcal{O}_{11}$. At the benchmark points $P_{25}$, $P_{26}$ and $P_{27}$ dark matter interacts with the nucleons through a linear combination of the interaction operators $\mathcal{O}_1$, $\mathcal{O}_3$, and $\mathcal{O}_{4}$. The operators $\mathcal{O}_1$ and $\mathcal{O}_4$ are the leading interaction operators in a velocity/momentum power series expansion. The operator $\mathcal{O}_3$ has been included in the linear combination defining the benchmark points $P_{25}$, $P_{26}$ and $P_{27}$, since it interferes with the leading operator $\mathcal{O}_1$.

\begin{table}
    \centering
    \begin{tabular}{cclcccc}
    \toprule
    Benchmark point   & $m_{\chi}$ (GeV) & $c_i^0\neq 0$  &  $N_1^{\rm th}$ & $N_1$ & $N_2^{\rm th}$& $N_2$  \\
    \midrule                                          
    $P_{1}$ & 10 & $m_v^2 c_3^0 = 20 $ & 1.7 & 3 &  86.3  & 81 \\
    $P_{2}$ & 10 & $m_v^2 c_5^0 = 120 $ &1.2 & 2 & 39.8 & 56  \\
    $P_{3}$ & 10 & $m_v^2 c_6^0 = 3000 $ & 2 & 3 & 75.9 & 71  \\
    $P_{4}$ & 10 & $m_v^2 c_7^0 = 2600 $ & 1.2 & 2 & 78 & 101 \\
    $P_{5}$ & 10 & $m_v^2 c_8^0 = 6$ & 1.4 & 2 & 87.3 & 82 \\
    $P_{6}$ & 10 & $m_v^2 c_9^0 = 80 $ & 1.8 & 3 & 73.8 & 96 \\
    $P_{7}$ & 10 & $m_v^2 c_{10}^0 = 50 $ & 1.7 & 3 & 69.9& 65  \\
    $P_{8}$ & 10 & $m_v^2 c_{11}^0 = 0.1$ & 1.5 & 2 & 48.9 & 67 \\
     \midrule  
    $P_{9}$ & 50 & $m_v^2 c_3^0 = 0.8 $ & 17.6 & 28 & 58.5 & 54 \\
    $P_{10}$ & 50& $m_v^2 c_5^0 = 12 $ & 34.4 & 31 & 75.5 & 98 \\
    $P_{11}$ & 50& $m_v^2 c_6^0 = 100 $ & 16.6 & 27 &  39.2 & 36 \\
    $P_{12}$ & 50 & $m_v^2 c_7^0= 300 $ & 17.9 & 29 & 52.8 & 49 \\
    $P_{13}$ & 50 & $m_v^2 c_8^0 = 0.5$ & 15.6 & 26 & 41.3 & 58 \\
    $P_{14}$ & 50 & $m_v^2 c_9^0 = 8$ &  24.8 & 38 & 59.1 & 79 \\
    $P_{15}$ & 50 & $m_v^ 2c_{10}^0 = 4$ & 20.1  & 32 & 60.2 & 56 \\
    $P_{16}$ & 50 & $m_v^2 c_{11}^0 = 0.01$ & 22.2 & 34 & 54.9 & 74 \\
     \midrule  
    $P_{17}$ & 250 & $m_v^2 c_3^0$ = 0.8 & 19.8 & 31 & 22.5 & 35 \\
    $P_{18}$ & 250& $m_v^2 c_5^0$ = 12 & 29.6 & 44 & 28.6 & 42 \\
    $P_{19}$ & 250& $m_v^2 c_6^0$ = 100 & 18.5 & 29 & 15.6 & 26 \\
    $P_{20}$ & 250 & $m_v^2 c_7^0$ = 300 & 8.7 & 10 & 17.3 & 28  \\
    $P_{21}$ & 250 & $m_v^2 c_8^0$ = 0.5 & 9.2 & 10 & 14.3 & 24 \\
    $P_{22}$ & 250 & $m_v^2 c_9^0$ = 8 &14.3 & 24 & 19.7 & 31 \\
    $P_{23}$ & 250 & $m_v^2 c_{10}^0$ = 4 & 15 & 25 & 21.6 & 34 \\   
    $P_{24}$ & 250 & $m_v^2 c_{11}^0$ = 0.01 & 15.6 & 26 & 19.1 & 30  \\
     \midrule  
    $P_{25}$ & 10 & $m_v^2 c_1^0= 1.5\times 10^{-3}$ &  1.3 & 2 & 39.2 & 36 \\
    			&	   &	 $m_v^2 c_4^0 = 1.5$ &   & & & \\
    			&       &   $m_v^2 c_3^0=15$ &   & & & \\
     \midrule  
    $P_{26}$ & 50 & $m_v^2 c_1^0= 2.5\times 10^{-4}$ & 16.4  & 27 & 46.8 & 43 \\
    			&	   &	 $m_v^2 c_4^0 = 0.25$ &   & & & \\
    			&       &   $m_v^2 c_3^0=0.7$ &   & & & \\
     \midrule  
    $P_{27}$ & 250 & $m_v^2 c_1^0= 5\times 10^{-4}$ &  34.9 & 50 &  55.6 & 75 \\
    			&	   &	 $m_v^2 c_4^0 = 0.5$ &   & & & \\
    			&       &   $m_v^2 c_3^0=1.2$ &   & & & \\
     \bottomrule
    \end{tabular}
    \caption{Properties of the benchmark points studied in this paper. For each benchmark point, this table shows the dark matter particle mass and the coupling constants different from zero. We also report $N_j^{\rm th}$ and $N_j$, respectively the expected and the ``observed'' number of dark matter scattering events in the Germanium ($j=1$) and Xenon ($j=2$) detectors.}
    \label{tab:benchmarks}
\end{table}

\section{Ton-scale direct detection experiments}
\label{sec:data}
We simulate our synthetic data from the benchmark points $P_1$ -- $P_{27}$. In the simulations, we assume the ton-scale Germanium and Xenon detectors described in the following. For each target material we include the most abundant stable isotopes present in Nature. For Germanium, the most abundant isotopes are $^{70}$Ge, $^{72}$Ge, $^{73}$Ge, $^{74}$Ge and $^{76}$Ge, whereas for Xenon, they are $^{128}$Xe, $^{129}$Xe, $^{130}$Xe, $^{131}$Xe, $^{132}$Xe, $^{134}$Xe, and $^{136}$Xe. We calculate the mass fractions $\xi_T$ of the 5 Germanium isotopes and of the 7 Xenon isotopes listed above from the relative abundances reported in Tab.~1 of Ref.~\cite{Feng:2011vu}. In general, different isotopes have distinct nuclear response functions. For the Germanium and the Xenon isotopes considered in the analyses, we calculate the nuclear response functions using our {\sffamily FORTRAN} version of the {\sffamily Mathematica} package described in Ref.~\cite{Anand:2013yka}. 

\subsection{Germanium}
For the ton-scale Germanium detector, we assume a Gaussian energy resolution with energy dispersion~\cite{Akrami:2010dn}
\begin{equation}
\sigma = \sqrt{\left(0.293\right)^2 + \left(0.056\right)^2 \left(E_R/ {\rm keV}\right)} \,,
\end{equation}
and a constant experimental efficiency $\mathcal{E}=0.3$. Accordingly, we calculate the differential rate of dark matter scattering events per unit time and unit Germanium detector mass as follows 
\begin{equation}
\frac{{\rm d}\mathcal{R}^{(1)}}{{\rm d}E_{\mathcal{O}}}=  \mathcal{E}\int_{0}^{\infty} dE_{R} \frac{1}{\sqrt{2\pi\sigma^2}} \exp\left[-\frac{(E_{R}-E_{\mathcal{O}})^2}{2\sigma^2}\right]  \frac{{\rm d}\mathcal{R}}{{\rm d}E_{R}} \,.
\label{eq:rate-Ge}
\end{equation}   
In Eq.~(\ref{eq:rate-Ge}), ${\rm d}\mathcal{R}/{\rm d}E_{R}$ is defined as in Eq.~(\ref{rate_theory}) and $E_{\mathcal{O}}$ is the observed energy. The latter coincides with the true nuclear recoil energy $E_{R}$ in the limit of infinite energy resolution. In our analyses we consider the time average of Eq.~(\ref{eq:rate-Ge}). The total number of scattering events in the signal region ($E_{\rm inf}$, $E_{\rm sup}$) is then given by
\begin{align}
\mu_S^{(1)}(m_\chi,\mathbf{c}) = MT\int_{E_{\rm inf}}^{E_{\rm sup}} \, \frac{{\rm d}\mathcal{R}}{{\rm d}E_{\mathcal{O}}} \, {\rm d}E_{\mathcal{O}} .
\label{eq:muS}
\end{align}
In our simulations we assume $E_{\rm inf}=10$~keV and $E_{\rm inf}=100$~keV, and a raw exposure for the Germanium detector of $MT=1000\times365$ kg-day. 

\subsection{Xenon}
The ton-scale Xenon detector considered in this paper has the same energy resolution and efficiency of the LUX experimental apparatus, but a larger exposure given by $MT=1000\times365$ kg-day. In the following, we briefly review how to calculate the expected number of dark matter scattering events in a detector with the same features as LUX.

The differential spectrum of the variable $S_1$, i.e. the observed number of dark matter induced photoelectrons (PE), is given by 
\begin{equation}
\frac{{\rm d}\mathcal{R}^{(2)}}{{\rm d}S_1} = \mathcal{E}(S_1)\sum_{n=1}^{+\infty} {\rm Gauss}(S_1 | n,\sqrt{n}\sigma_{\rm PMT})  \int_{0}^{\infty}{\rm d}E_{R} \,{\rm Poiss}(n|\nu(E_{R})) \frac{{\rm d}\mathcal{R}}{{\rm d}E_{R}} \,.
\label{eq:rate_Xe}
\end{equation}
In Eq.~(\ref{eq:rate_Xe}), the Gaussian of mean $n$ and standard deviation $\sqrt{n}\sigma_{\rm PMT}$, with $\sigma_{\rm PMT}=0.37$, gives the probability of observing $S_1$ PE when the true number of dark matter induced PE is $n$. In the same equation, the Poisson distribution of mean $\nu(E_R)$ gives the probability of producing $n$ PE from a recoil energy $E_R$. In the case of LUX, the function $\nu(E_R)$ can be extracted from the panel (b) of  Fig.~3 in Ref.~\cite{Akerib:2013tjd}. In our simulations we assume the same expression for $\nu(E_R)$, and the experimental efficiency reported in Fig.~9 of Ref.~\cite{Akerib:2013tjd}, multiplied by an additional factor 1/2, corresponding to the 50\% nuclear recoil acceptance quoted by the LUX collaboration. In our analyses, we  consider the time average of Eq.~(\ref{eq:rate_Xe}). The differential rate ${\rm d}\mathcal{R}/{\rm d}E_{R}$ appearing in Eq.~(\ref{eq:rate_Xe}) is defined as in Eq.~(\ref{rate_theory}), and the total number of scattering events in the signal region ($S_{1}^{\rm inf}$, $S_{1}^{ \rm sup}$) is given by
\begin{align}
\mu^{(2)}_S(m_\chi,\mathbf{c}) = MT\int_{S_{1}^{\rm inf}}^{S_{1}^{\rm sup}} \, \frac{{\rm d}\mathcal{R}}{{\rm d}S_{1}} \,{\rm d}S_1 .
\label{eq:muS}
\end{align}
In the simulations, we assume $S_{1}^{\rm inf}=2$~PE and $S_{1}^{\rm sup}=30$~PE.

\subsection{Background}
Future ton-scale detectors will measure recoil events originating from local radioactivity, cosmic rays, and other experimental backgrounds, besides nuclear recoil events induced by dark matter scattering in the target material. 

For the irreducible background events in the ton-scale Germanium and Xenon detectors, we assume the following energy spectrum
\begin{equation}
\frac{{\rm d}\mathcal{R}^{(j)}_{\rm B}}{{\rm d}\hat{E}_j} = \frac{\eta}{b_j-a_j} +  \frac{\eta}{\epsilon_j\left[ \exp(-a_j/\epsilon_j)-\exp(-b_j/\epsilon_j)\right]} e^{-\hat{E}_j/\epsilon_j} \,,
\label{eq:background}
\end{equation}
where $a_j$ ($b_j$) is the lower (upper) limit of the signal region. We have introduced an index $j$ to characterize the quantities depending on the detector type. $j=1$ refers to the Germanium detector, whereas $j=2$ identifies the Xenon detector. With this notation, $\hat{E}_1=E_\mathcal{O}$ and  $\hat{E}_2=S_1$. In Eq.~(\ref{eq:background}), $a_1=10$ keV, $b_1=100$ keV and $\epsilon_1=10$ keV, whereas $a_2=2$ PE, $b_2=30$ PE and $\epsilon_2=2$ PE. The energy spectrum in  Eq.~(\ref{eq:background}) includes a flat component and an exponentially decreasing component, as expected for dark matter direct detection experiments~\cite{Akrami:2010dn}. In Eq.~(\ref{eq:background}) $\eta=0.5$, which implies one background event in the signal region, both for the Germanium detector and for the Xenon detector. Accordingly, we define the quantity 
 \begin{equation}
\mu_{\rm tot}^{(j)}(m_{\chi},{\mathbf{c}}) = \mu^{(j)}_{S}(m_{\chi},{\mathbf{c}}) + 1 
\end{equation}
as the total number of expected events in the signal region. Again, $j=1$ refers to the Germanium detector, and $j=2$ corresponds to the Xenon detector. 

\subsection{Synthetic data}
\label{sec:simulation}
We now describe the procedure that we have followed to simulate synthetic data given a benchmark point. We illustrate the procedure for a generic detector of type $j$. A sample of synthetic direct detection data is a set of $N_{j}$ recoil energies, or PE, $\{\hat{E}_j\}_{i=1,\dots,N_j}$. The datapoints of this set are randomly generated from the recoil energy spectrum of a benchmark point (e.g. one of the points $P_{1}$ -- $P_{27}$). 

More specifically, the simulation of synthetic direct detection data consists of two parts: (1) generating the number $N_j$; (2) sampling the events $\{\hat{E}_j\}_{i=1,\dots,N_j}$. We randomly sample the number $N_j$ of observed events from a Poisson distribution of mean $\mu_{\rm tot}(\hat{m}_\chi,\hat{\mathbf{c}})$, where the parameters $(\hat{m}_\chi,\hat{\mathbf{c}})$ identify one of the benchmark points $P_{1}$ -- $P_{27}$. Subsequently, we randomly sample $N_j$ recoil energies, or PE, from the spectral function 
\begin{equation}
\hat{f}^{(j)}(\hat{E}_{j}) \equiv f^{(j)}(\hat{E}_j,\hat{m}_\chi,\hat{\mathbf{c}}) 
\end{equation}
with
\begin{equation}
f^{(j)}(\hat{E}_{j},m_\chi,\mathbf{c}) = \frac{1}{\mu^{(j)}_{\rm tot}(m_{\chi},{\mathbf{c}})} \left[ \frac{{\rm d}\mathcal{R}^{(j)}}{{\rm d}\hat{E}_{j}} \left( \hat{E} _{j},m_\chi,\mathbf{c} \right)  + \frac{{\rm d}\mathcal{R}^{(j)}_{\rm B}}{{\rm d}\hat{E}_j} \left( \hat{E} _{j} \right)  \right]\,.
\label{eq:spectralf}
\end{equation}

In our analyses, we have repeated this procedure twice for each benchmark point $P_{1}$ -- $P_{27}$. Once assuming the Germanium detector ($j=1$) and once assuming the Xenon detector ($j=2$). Tab.~\ref{tab:benchmarks} reports the number of simulated events for each benchmark point and for each target material.

\section{Statistical framework}
\label{sec:statistics}
We now introduce the statistical methods applied to the analysis of our synthetic direct detection data. The aim of our analyses is to reconstruct $m_\chi$ and the coupling constants $c_i^0$ from about 100 scattering events randomly generated from the benchmark points $P_{1}$ -- $P_{27}$.

As explained in Sec.~\ref{sec:simulation}, for each benchmark point we generate two samples of synthetic data. One sample for the ton-scale Germanium detector ($j=1$) and one sample for the ton-scale Xenon detector ($j=2$). Each sample consists of a $(N_j+1)$-dimensional array of datapoints, $\mathbf{d}_{j}\equiv (N_j,\{\hat{E}_j\}_{i=1,\dots,N_j})$. To each individual dataset $\mathbf{d}_{j}$ we assign the Likelihood function 
\begin{eqnarray}
-\ln \mathcal{L}^{(j)}(\mathbf{d}_{j}, |m_\chi,\mathbf{c}) &=& \mu^{(j)}_{\rm tot}(m_\chi,\mathbf{c}) - N_j \ln [\mu^{(j)}_{\rm tot}(m_\chi,\mathbf{c})]  \nonumber\\
&+& \sum_{i=1}^{N_j}\log \frac{f^{(j)}(\hat{E}_j,\hat{m}_\chi,\hat{\mathbf{c}})}{f^{(j)}(\hat{E}_{j},m_\chi,\mathbf{c})}\,.
\label{eq:like}
\end{eqnarray}
Accordingly, the total Likelihood function takes the following form
\begin{equation}
-\ln \mathcal{L}_{\rm tot}(\mathbf{d}_1,  \mathbf{d}_2, |m_\chi,\mathbf{c}) =  -\sum_{j=1,2} \ln \mathcal{L}^{(j)}(\mathbf{d}_j, |m_\chi,\mathbf{c}) \,.
\label{eq:like_tot}
\end{equation}
In Eq.~(\ref{eq:like}) the first two terms constrain $\mu^{(j)}_{\rm tot}(m_\chi,\mathbf{c})$ to match $N_j$, whereas the sum in the second line contains the information on the spectrum of the observed events. The spectral information is encoded in the functions $f^{(j)}$ defined in Eq.~(\ref{eq:spectralf}).

We use the Likelihood function in Eq.~(\ref{eq:like_tot}) to construct the posterior probability density function (PDF) of the model parameters (a Bayesian appraoch) and their profile Likelihood (a frequentist approach). The posterior PDF, $\mathcal{P}(m_\chi,\mathbf{c} | \mathbf{d}_1,  \mathbf{d}_2)$, is related to the total Likelihood function by Bayes' theorem, according which 
\begin{equation}
\mathcal{P}(m_\chi,\mathbf{c} | \mathbf{d}_1,  \mathbf{d}_2) = \frac{\mathcal{L}_{\rm tot}(\mathbf{d}_1,  \mathbf{d}_2, |m_\chi,\mathbf{c}) \pi(m_\chi,\mathbf{c})}{\mathcal{E}(\mathbf{d}_1,  \mathbf{d}_2)}\,.
\label{eq:bayes}
\end{equation} 
In Eq.~(\ref{eq:bayes}), $\pi(m_\chi,\mathbf{c})$ is the prior PDF which encodes our prejudice on the model parameter before having seen the data. The Bayesain evidence $\mathcal{E}(\mathbf{d}_1,  \mathbf{d}_2)$ is independent of the model parameters and it hence plays the role of a normalization constant when performing parameter inference, as in the present analysis. Analyzing our synthetic direct detection data, we assume log-priors both for the dark matter mass and for the coupling constants. Log-priors allow to sample the posterior PDF varying the model parameters within prior ranges spanning several orders of magnitude. Tab.~\ref{tab:priors} shows the prior ranges assumed in our analyses.

\begin{table}
    \centering
    \begin{tabular}{lclc}
    \toprule
    Parameter         & Type & Prior range &  Prior type \\
    \midrule                                          
    $\log_{10} (m_v^2 c_1^\tau)$ & model parameter & $[-5,1]$ & log-prior  \\
    $\log_{10} (m_v^2c_3^\tau)$ & model parameter & $[-4,4]$ & log-prior  \\
    $\log_{10} (m_v^2c_4^\tau)$ & model parameter & $[-2,3]$ & log-prior  \\
    $\log_{10} (m_v^2c_5^\tau)$ & model parameter & $[-4,4]$ & log-prior  \\
    $\log_{10} (m_v^2c_6^\tau)$ & model parameter & $[-4,4]$ & log-prior  \\
    $\log_{10} (m_v^2c_7^\tau)$ & model parameter & $[-4,4]$ & log-prior  \\
    $\log_{10} (m_v^2c_8^\tau)$ & model parameter & $[-4,4]$ & log-prior  \\
    $\log_{10} (m_v^2c_9^\tau)$ & model parameter & $[-4,4]$ & log-prior  \\
    $\log_{10} (m_v^2c_{10}^\tau)$ & model parameter & $[-4,4]$ & log-prior  \\
    $\log_{10} (m_v^2c_{11}^\tau)$ & model parameter & $[-4,4]$ & log-prior  \\
    $\log_{10} (m_{\chi}/{\rm GeV})$ & model parameter & $[0.1,3 (4)]$  & log-prior  \\
    \bottomrule
    \end{tabular}
    \caption{List of model parameters. For each parameter we report the type of assumed prior PDF, and the corresponding prior range. For the benchmark points with $m_\chi=250$~GeV, we extend the dark matter prior range to $[0.1,4]$. Following~\cite{Anand:2013yka}, we express the coupling constants in units of $m_v^{-2}=(246.2~{\rm GeV})^{-2}$. }
    \label{tab:priors}
\end{table}

We present our results in terms of 2D marginal poster PDFs and 2D profile likelihoods. The 2D marginal posterior PDF of the model parameters $m_\chi$-$c_1^0$, for instance,  is defined as follows
\begin{equation}
\mathcal{P}_{\rm marg}(m_\chi, c_1^0|\mathbf{d}_1,\mathbf{d}_2)  \propto \int d c_{3}^0 \dots d c_{11}^0\, \mathcal{P}(m_\chi,\mathbf{c}|\mathbf{d}_1,\mathbf{d}_2) \,,
\label{eq:marg}
\end{equation} 
whereas the 2D profile likelihood of the same model parameters is given by
\begin{equation}
\mathcal{L}_{\rm prof}(\mathbf{d}_1,\mathbf{d}_2|m_\chi,c^0_1) \propto \max_{c_3^0,\dots,c_{11}^0}  \mathcal{L}_{\rm tot}(\mathbf{d}_1,  \mathbf{d}_2, |m_\chi,\mathbf{c})\,.
\label{eq:prof_likelihood}
\end{equation} 
Similar expressions holds for the other pairs of model parameters.

The integral in Eq.~(\ref{eq:marg}) is dominated by the tails of the posterior PDF, when the latter extends over a large volume  in parameter space compared to the region where the Likelihood peaks. Highly tailed posterior PDFs are generated by prior PDFs containing more information than the Likelihood function. In this case, regions in parameters space where the Likelihood is small can be overweighted by the marginalization procedure. Profile likelihoods are statistical indicators insensitive to these ``volume effects'', though they are computationally demanding quantities, contrary to marginal posterior PDFs. Marginal posterior PDFs and profile likelihoods are therefore complementary statistical indicators in the context of the dark matter direct detection. 

Within the Bayesian approach to data analysis, limits on the coupling constants $c_i^{\tau}$ and on the mass $m_\chi$ are expressed in terms of $x$\% credible regions (CR). These regions of the parameter space contain $x$\% of the total posterior probability, and are such that $\mathcal{P}_{\rm marg}$ at any point inside the region is larger than at any point outside the region. Within the frequentist approach to data anlysis, one can use the profile likelihood to construct approximate frequentist confidence intervals from an effective chi-square defined as $\Delta \chi^2_{\rm eff}~\equiv~-2 \ln \mathcal{L}_{\rm prof}/ \mathcal{L}_{\rm max}$, where $\mathcal{L}_{\rm max}$ is the absolute maximum of the Likelihood function. Under certain regularity conditions the distribution of $\Delta \chi^2_{\rm eff}$ converges to a chi-square distribution with a number of degrees of freedom equal to the number of relevant parameters (Wilks' theorem~\cite{2011JHEP...06..042F}), e.g. 2 for a 2D profile likelihood. 

We compute the posterior mean of the model parameters taking their average over the posterior PDF. We maximize the likelihood function in order to obtain the best fit points of the model parameters. Ideally, posterior means and best fit points should coincide in the limit of gaussian likelihood function. A difference between the two statistical indicators denotes a departure from the gaussian limit, or a dependence of the results on the choice of priors.

We use confidence intervals and credible regions to assess whether the mass $m_\chi$ and the couplings $c_j^{0}$ characterizing the benchmark points $P_1$--$P_{27}$ can be extracted from our synthetic direct detection data.

Finally, to sample the multidimensional Likelihood function in Eq.~(\ref{eq:like_tot}) we use the {\sffamily Multinest} program~\cite{Feroz:2008xx,Feroz:2007kg,Feroz:2013hea}. We use our own routines to evaluate event rates and the Likelihood function. Figures have been produced using the programs {\sffamily GetDist}~\cite{Lewis:2002ah},  {\sffamily Getplots}~\cite{Austri:2006pe} and {\sffamily Matlab}. 

\section{Results}
\label{sec:results}
We now analyze the synthetic data generated from the benchmark points $P_1$ -- $P_{27}$ as explained in Sec.~\ref{sec:simulation}. We tackle this problem within the statistical framework of Sec.~\ref{sec:statistics}. First, we focus on the benchmark points $P_1$ -- $P_{24}$, where the dark matter-nucleon interaction is described by a single operator (Sec.~\ref{sec:P1P24}). Then, we analyze the case in which the dark matter particle interacts with the detector nucleons through a linear combination of non-relativistic operators (Sec.~\ref{sec:P25P27}). Our goal is to assess whether the benchmark values of $m_\chi$ and of the constants $c_i^0\neq 0$ at $P_1$ -- $P_{27}$ can be extracted from our synthetic data.

\subsection{Prospects for benchmark points $P_1$ -- $P_{24}$}
\label{sec:P1P24}
\begin{figure}[t]
\begin{center}
\includegraphics[width=\textwidth]{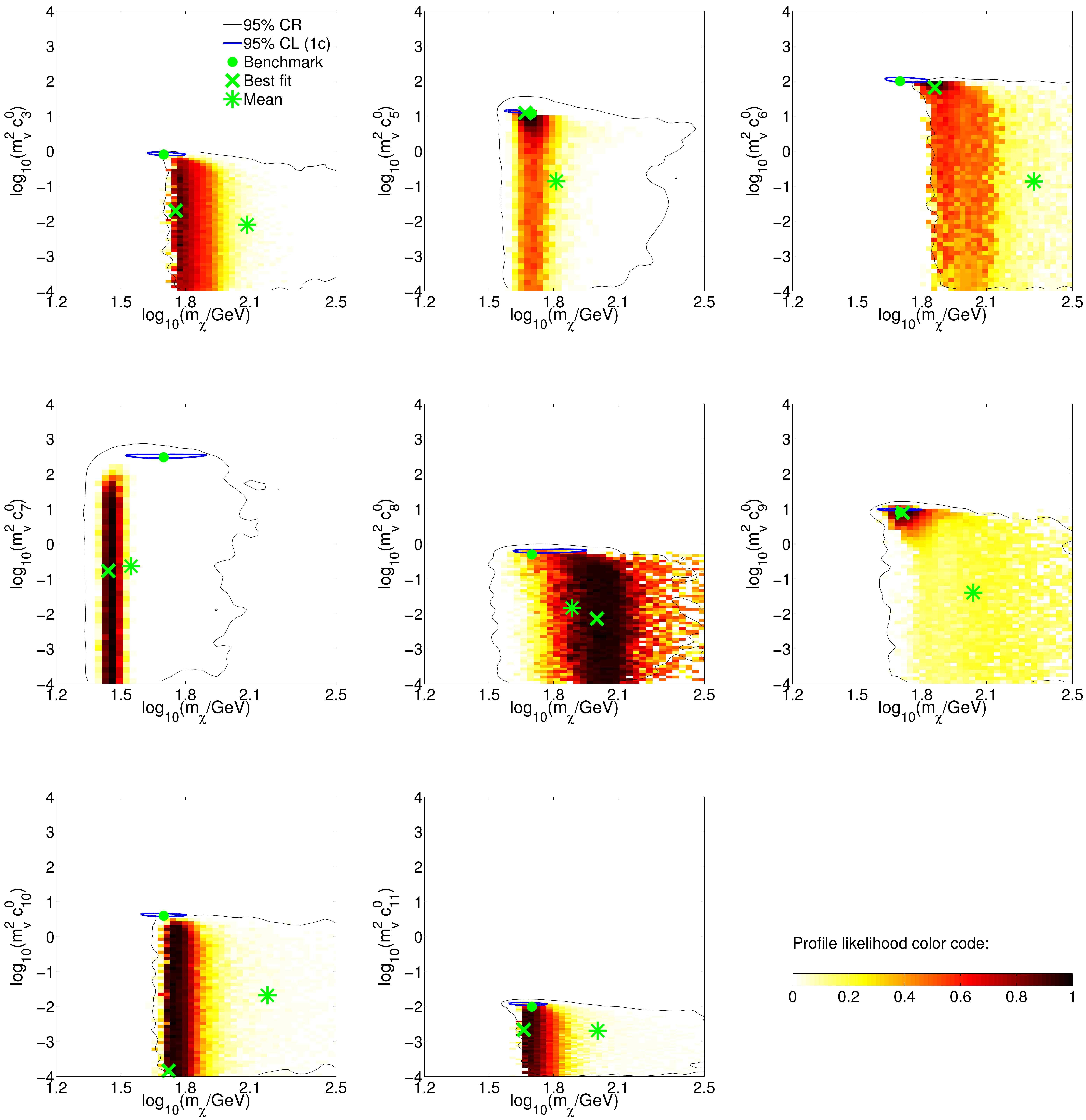}
\end{center}	
\caption{Analysis of the synthetic data randomly generated from the benchmark points $P_9$ -- $P_{16}$. These benchmark points are characterized by $m_\chi=50$~GeV and by the coupling constants listed in Tab.~\ref{tab:benchmarks}. Different panels refer to distinct benchmark points. In all panels, results are presented in terms of 2D profile likelihoods (colored regions), 95\% credible regions (black contours) and 95\% confidence levels (blue contours). For each benchmark point, we have constructed the 2D profile likelihoods and credible regions, by fitting the full 11-dimensional effective theory of Sec.~\ref{sec:theory} to our simulated data (fitting procedure B). In each panel, we have derived the 95\% confidence levels, by fitting $m_\chi$ and the coupling constant on the y-axis to our simulated data (fitting procedure A). Green circles, crosses and stars are the benchmark points, the best fit values and the means resulting from the fitting procedure B. We find that for the benchmark points $P_{10}$, $P_{11}$ and $P_{14}$, corresponding to the interaction operators $\mathcal{O}_5$, $\mathcal{O}_6$ and $\mathcal{O}_9$, the leading dark matter-nucleon interaction and the dark matter mass can be extracted from the synthetic data directly, without any further assumption.}
\label{fig:P9P16}
\end{figure}

Tab.~\ref{tab:benchmarks} summarizes the properties of the benchmark points $P_1$ -- $P_{24}$. To analyze the synthetic data generated from the points $P_1$ -- $P_{24}$, we adopt two strategies:
\begin{enumerate}
\item {\it Fitting procedure A}. For each benchmark point $P_A$, with $1 \le A \le 24 $, we fit the dark matter-nucleon effective theory of Sec.~\ref{sec:theory} to our synthetic data, assuming in the fit 2 free parameters only. The 2 free parameters are $m_\chi$ and $c^0_k\neq 0$, where $c^0_k$ is the only coupling constant different from zero at $P_A$. When analyzing our synthetic data following the fitting procedure A, we fix to zero the coupling constants $c_j^0$ with $j\neq k$. The fitting procedure A is the one commonly used in fitting the leading spin-independent and spin-dependent interactions, i.e. $\mathcal{O}_1$ and $\mathcal{O}_4$, to direct detection data. The blue lines in Figs.~\ref{fig:P9P16}, \ref{fig:P1P8} and \ref{fig:P17P24} represent the 95\% CL contours found applying the fitting procedure A to the benchmark points $P_1$ -- $P_{24}$.
\item {\it Fitting procedure B}. Alternatively, we fit the dark matter-nucleon effective theory of Sec.~\ref{sec:theory} to our synthetic data assuming in the fit 11 free parameters. The 11 free parameters are $m_\chi$ and the 10 coupling constants $c_i^0$, with $i=1,3,\dots,11$. In contrast to the fitting procedure A, the fitting procedure B extracts the correct dark matter-nucleon interaction type from the data directly. It requires a comprehensive exploration of the full 11-dimensional parameter space of the effective theory defined in Sec.~\ref{sec:theory}. The black lines and the colored regions in Figs.~\ref{fig:P9P16}, \ref{fig:P1P8} and \ref{fig:P17P24} represent, respectively, the 95\% CR contours and the 2D profile likelihoods that we obtain applying the fitting procedure B to the benchmark points $P_1$ -- $P_{24}$.
\end{enumerate}
Fig.~\ref{fig:P9P16} summarizes our findings for the benchmark points $P_9$ -- $P_{16}$. We present our results in the 8 planes $m_\chi$-$c_i^0$, with $i=3,5\dots,11$. Applying the fitting procedure A to the synthetic data generated from the points $P_9$ -- $P_{16}$, we produce the 95\% CL contours (blue lines) shown in Fig.~\ref{fig:P9P16}. These contours are relatively narrow, compared to the large volume of the parameter space explored in our analyses. In addition, in each panel of Fig.~\ref{fig:P9P16} the benchmark points (green circles) are fully contained in (or at the very edge of) the blue contours. We conclude that $m_\chi$ and the constant $c_{k}^0\neq 0$ can be extracted from the synthetic data, applying the fitting procedure A to the benchmark points $P_9$ -- $P_{16}$. 

\begin{figure}[t]
\begin{center}
\begin{minipage}[t]{0.49\linewidth}
\centering
\includegraphics[width=\textwidth]{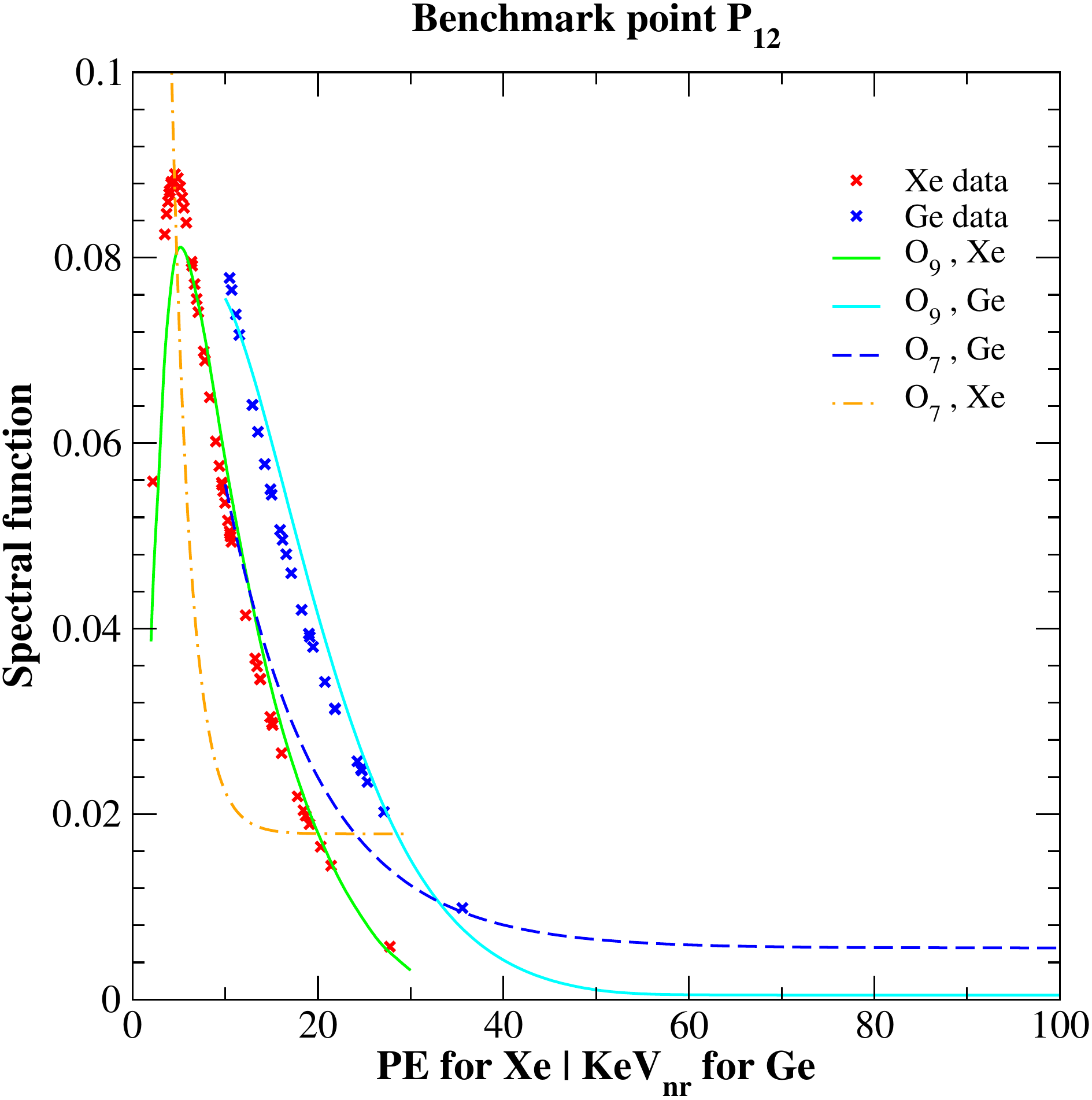}
\end{minipage}
\begin{minipage}[t]{0.49\linewidth}
\centering
\includegraphics[width=\textwidth]{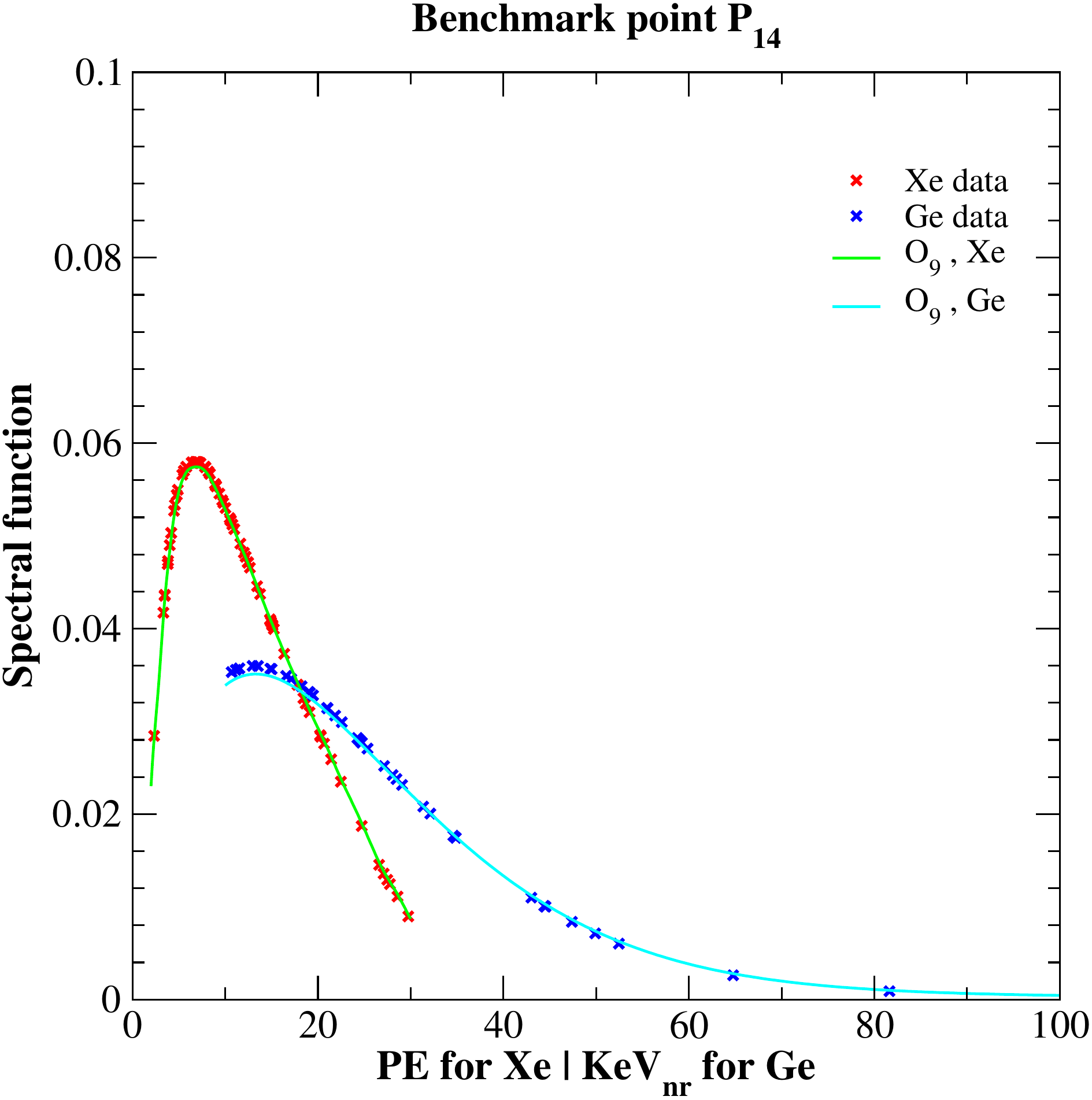}
\end{minipage}
\end{center}
\caption{Compared analysis of the benchmark points $P_{12}$ (left panel) and $P_{14}$ (right panel). For each benchmark point, we report the values of the associated spectral functions evaluated at the simulated recoil energies for the Germanium detector (blue crosses), and at the simulated PE for the Xenon detector (red crosses). In the legends, we use the symbol ``$\mathcal{O}_k$'' to indicate that the corresponding line has been obtained setting $c_{i}^0=0$, for $i\neq k$, and $c_{k}^0$ at the best fit value of the fitting procedure B. At $P_{14}$ the benchmark spectral functions, and the lines denoted by $\mathcal{O}_9$ are almost identical (i.e. the original benchmark point is well reconstructed). At $P_{12}$, instead, the benchmark spectral functions, and the lines denoted by $\mathcal{O}_7$ differ significantly. In contrast, at $P_{12}$ the benchmark spectral functions match the the lines denoted by $\mathcal{O}_9$ fairly well (i.e. the operator $\mathcal{O}_9$ is favored by the fit, though the data have been generated from $\mathcal{O}_7$). This figure illustrates why at $P_{12}$ the fitting procedure B performs less accurately than at $P_{14}$.}
\label{fig:P12P14}
\end{figure}

\begin{figure}[t]
\begin{center}
\includegraphics[width=\textwidth]{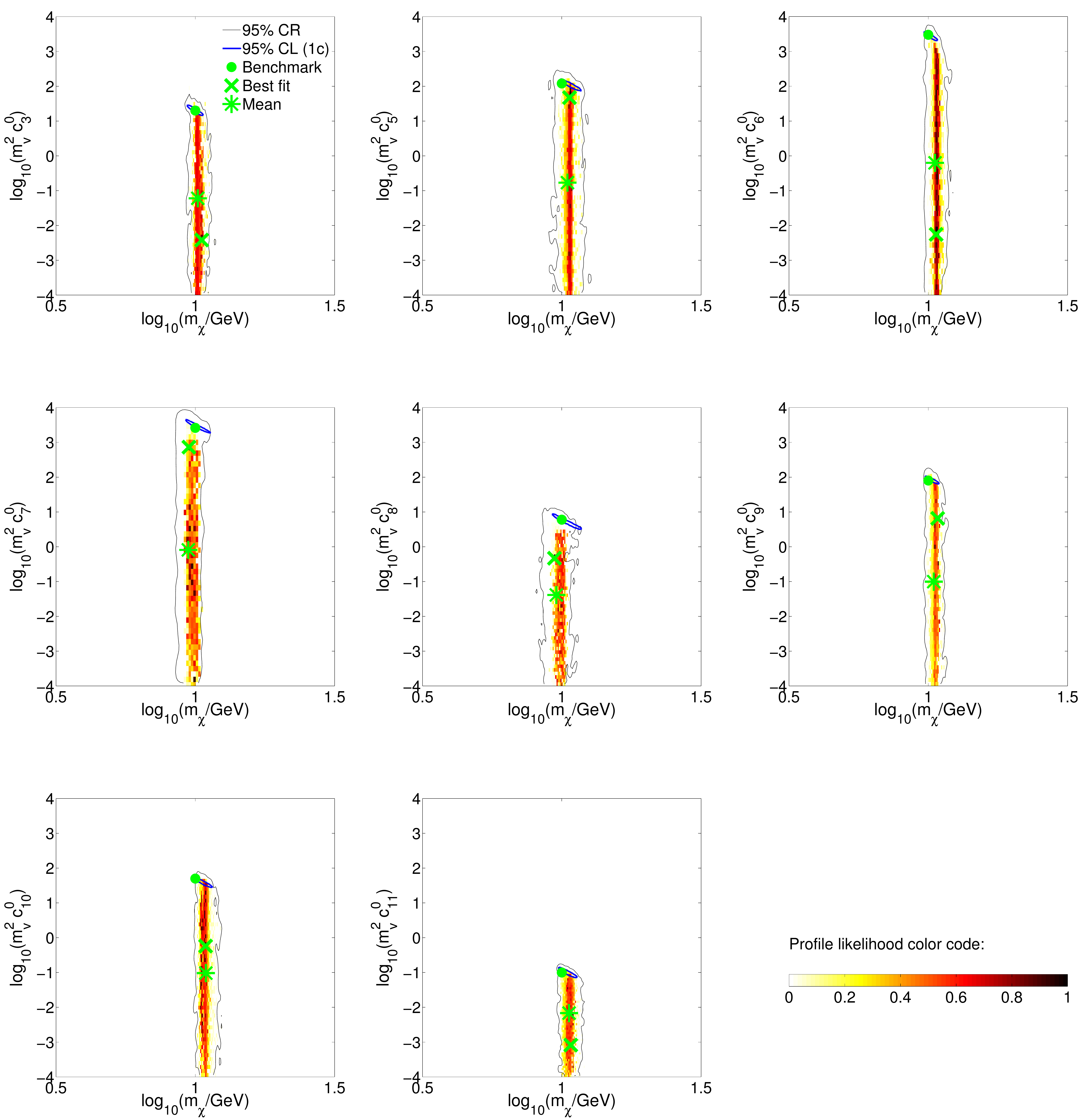}
\end{center}
\caption{Same as Fig.~\ref{fig:P9P16}, but for the benchmark points $P_1$ -- $P_{8}$. These benchmark points are characterized by $m_\chi=10$~GeV and by the coupling constants listed in Tab.~\ref{tab:benchmarks}.  } 
\label{fig:P1P8}
\end{figure}
\begin{figure}[t]
\begin{center}
\includegraphics[width=\textwidth]{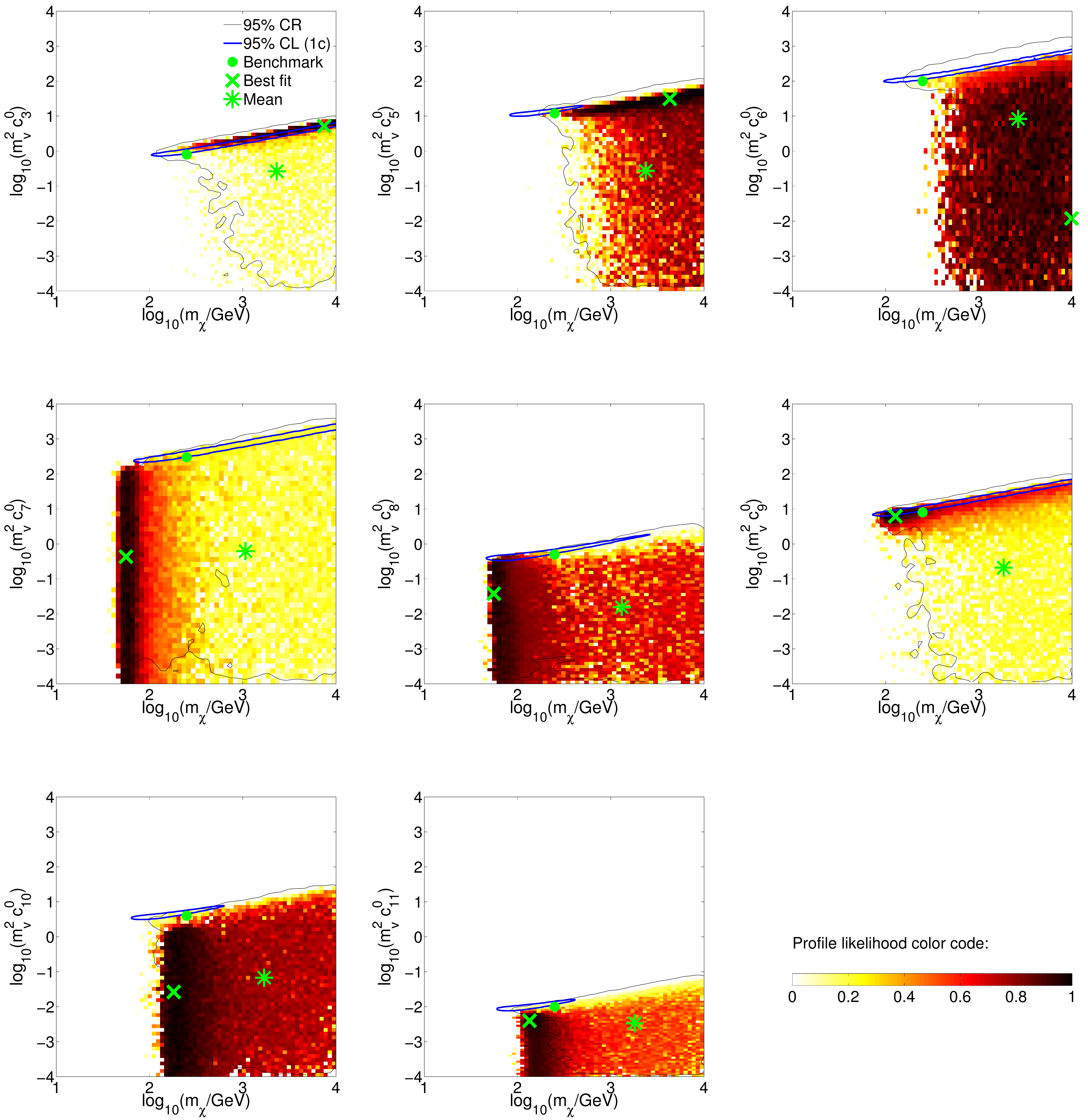}
\end{center}
\caption{Same as Fig.~\ref{fig:P9P16}, but for the benchmark points $P_{17}$ -- $P_{24}$. These benchmark points are characterized by $m_\chi=250$~GeV and by the coupling constants listed in Tab.~\ref{tab:benchmarks}. }
\label{fig:P17P24}
\end{figure}

However, the fitting procedure A can be only used when the form of the dark matter-nucleon interaction is known a priori. Though this is a reasonable simplifying assumption in a preliminary exploration of the data, it is far from being supported by any empirical evidence. In general, we have therefore to tackle a problem of more complex nature: extracting the correct dark matter-nucleon interaction type from the data directly. The fitting procedure B is the correct way of approaching this challenging problem.

We therefore apply the fitting procedure B to the synthetic data generated from the points $P_9$ -- $P_{16}$. Our aim is to assess whether extracting the correct dark matter-nucleon interaction type from about 100 scattering events is indeed feasible. From an analysis of the benchmark points $P_9$ -- $P_{16}$ based on the fitting procedure B, we extract the 95\% CR contours and the profile likelihoods shown in the 8 planes $m_\chi$-$c_i^0$, $i=3,5\dots,11$, of Fig.~\ref{fig:P9P16}. Importantly, we find that for the benchmark points $P_{10}$, $P_{11}$ and $P_{14}$, corresponding to the interaction operators $\mathcal{O}_5 = -i\vec{S}_\chi\cdot(\vec{q}/m_N\times\vec{v}^{\perp}_{\chi N})$, $\mathcal{O}_6 = (\vec{S}_\chi\cdot\vec{q}/m_N)(\vec{S}_N\cdot\vec{q}/m_N)$ and $\mathcal{O}_9 = -i\vec{S}_\chi\cdot(\vec{S}_N\times\vec{q}/m_N)$, the correct dark matter-nucleon interaction type and the dark matter mass can be extracted from the synthetic data directly. In fact, the best fit dark matter mass and coupling constants that we find applying the fitting procedure B to the benchmark points $P_{10}$, $P_{11}$ and $P_{14}$ match the corresponding benchmark values within a very good accuracy (with the exception of the dark matter mass at $P_{11}$). For instance, for the benchmark point $P_{14}$, we find the best fit values: $\log_{10}(m_\chi/{\rm GeV})=1.71$, $\log_{10}(m_v^2 c_9^0)=0.89$ and all the other coupling constants are compatible with zero. Remarkably, the values of $m_\chi$ and $c_9^0$ at the benchmark point $P_{14}$ are: $\log_{10}(m_\chi/{\rm GeV})=1.70$, $\log_{10}(m_v^2 c_9^0)=0.90$, and $c_i^0=0$, for $i\neq 9$. 

When applied to the  benchmark points $P_{9}$, $P_{12}$, $P_{13}$, $P_{15}$ and $P_{16}$, the fitting procedure B gives less accurate results. More specifically, from the synthetic data generated from these benchmark points, we are not able to extract the correct dark matter-nucleon interaction type. However, for the benchmark points  $P_{9}$, $P_{15}$ and $P_{16}$, we can accurately reconstruct the dark matter mass from an analysis of the corresponding synthetic data (see Tab.~\ref{tab:results} for a list of best fit values). From this study, the benchmark point $P_{12}$ appears as the most difficult to analyze.

At the benchmark points $P_{9}$, $P_{12}$, $P_{13}$, $P_{15}$ and $P_{16}$, the operator that better fits the synthetic data is not the operator from which the data have actually been generated. At the best fit point found applying the fitting procedure B to $P_{12}$, for instance, the operator $\mathcal{O}_9$ generates the majority of the expected recoil events, though at $P_{12}$ the data have been generated from the operator $\mathcal{O}_7$, with $c_{i}^0=0$ for $i\neq7$. Fig.~\ref{fig:P12P14} illustrates this result in a compared analysis of the benchmark points $P_{12}$ and $P_{14}$. 

Figs.~\ref{fig:P1P8} and \ref{fig:P17P24} summarize our findings for the benchmark points $P_1$ -- $P_8$ and $P_{17}$ -- $P_{24}$, respectively. Applying the fitting procedure A to the benchmark points $P_1$ -- $P_8$, we are always able to extract $m_\chi$ and $c_k^0\neq 0$ from the synthetic data. For the benchmark points $P_{17}$~--~$P_{24}$, we find that the fitting procedure A can only constrain the ratio $(c_k^0)^2/m_\chi$, since $v_{\rm min}$ is approximately independent of $m_{\chi}$, when $m_{\chi}$ is significantly larger than the target nuclei mass. For this reason the 95\% CL contours extracted from the fitting procedure A applied to $P_{17}$ -- $P_{24}$ are broader than in Figs.~\ref{fig:P9P16} and \ref{fig:P1P8}.

\begin{figure}[t]
\begin{center}
\includegraphics[width=\textwidth]{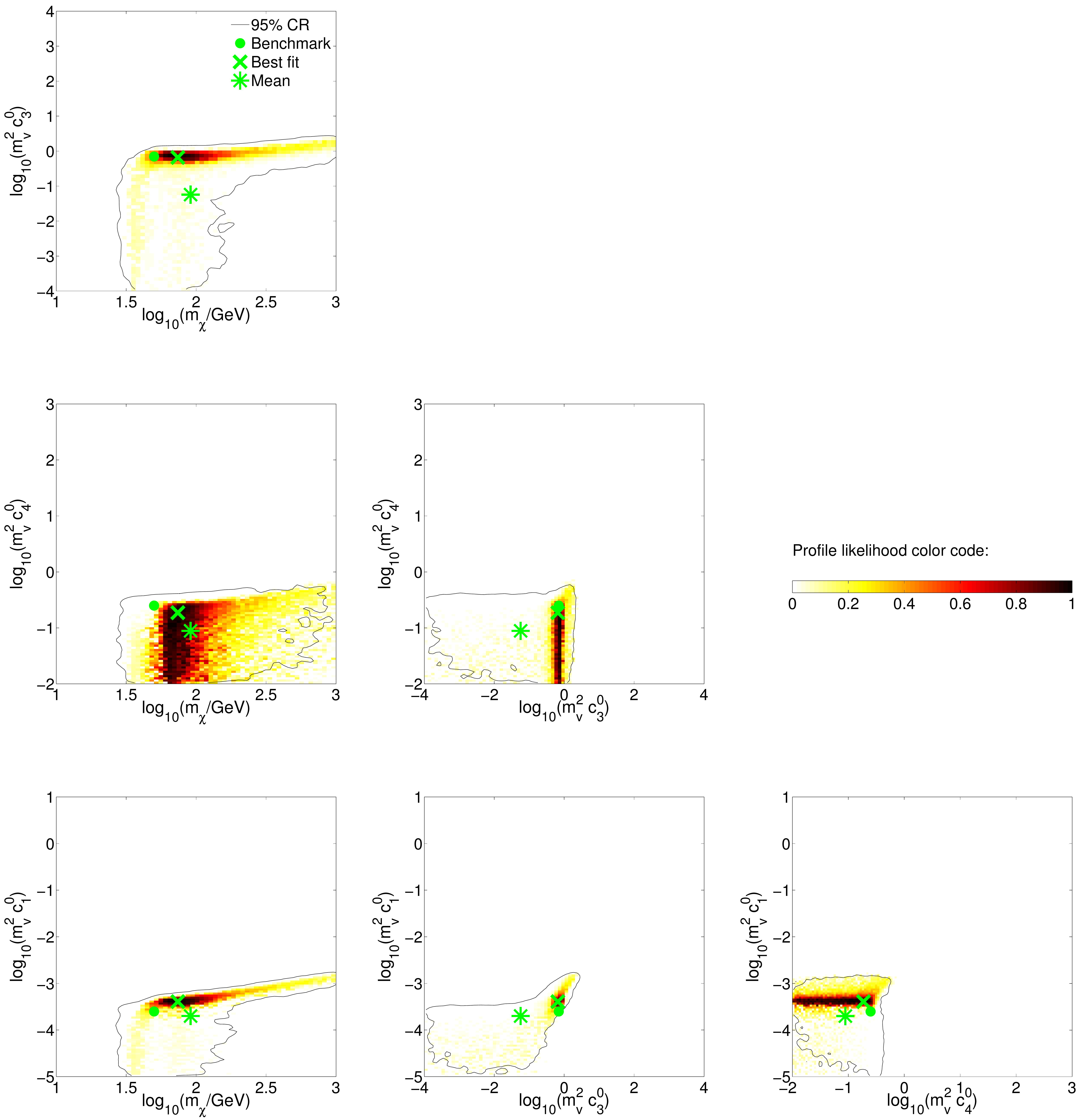}
\end{center}
\caption{Analysis of the synthetic data randomly generated from the benchmark point $P_{26}$. This benchmark point is characterized by $m_\chi=50$~GeV and by the coupling constants listed in Tab.~\ref{tab:benchmarks}. Different panels refer to distinct pairs of model parameters. In all panels, results are presented in terms of 2D profile likelihoods (colored regions) and 95\% credible regions (black contours). We have constructed the 2D profile likelihoods and credible regions, by fitting the full 11-dimensional effective theory of Sec.~\ref{sec:theory} to our simulated data. For this benchmark point, we find that the dark matter mass and coupling constants $c_1^0$,  $c_3^0$ and $c_4^0$ can be simultaneously extracted from our synthetic data.}
\label{fig:P26}
\end{figure}
\begin{figure}[t]
\begin{center}
\includegraphics[width=\textwidth]{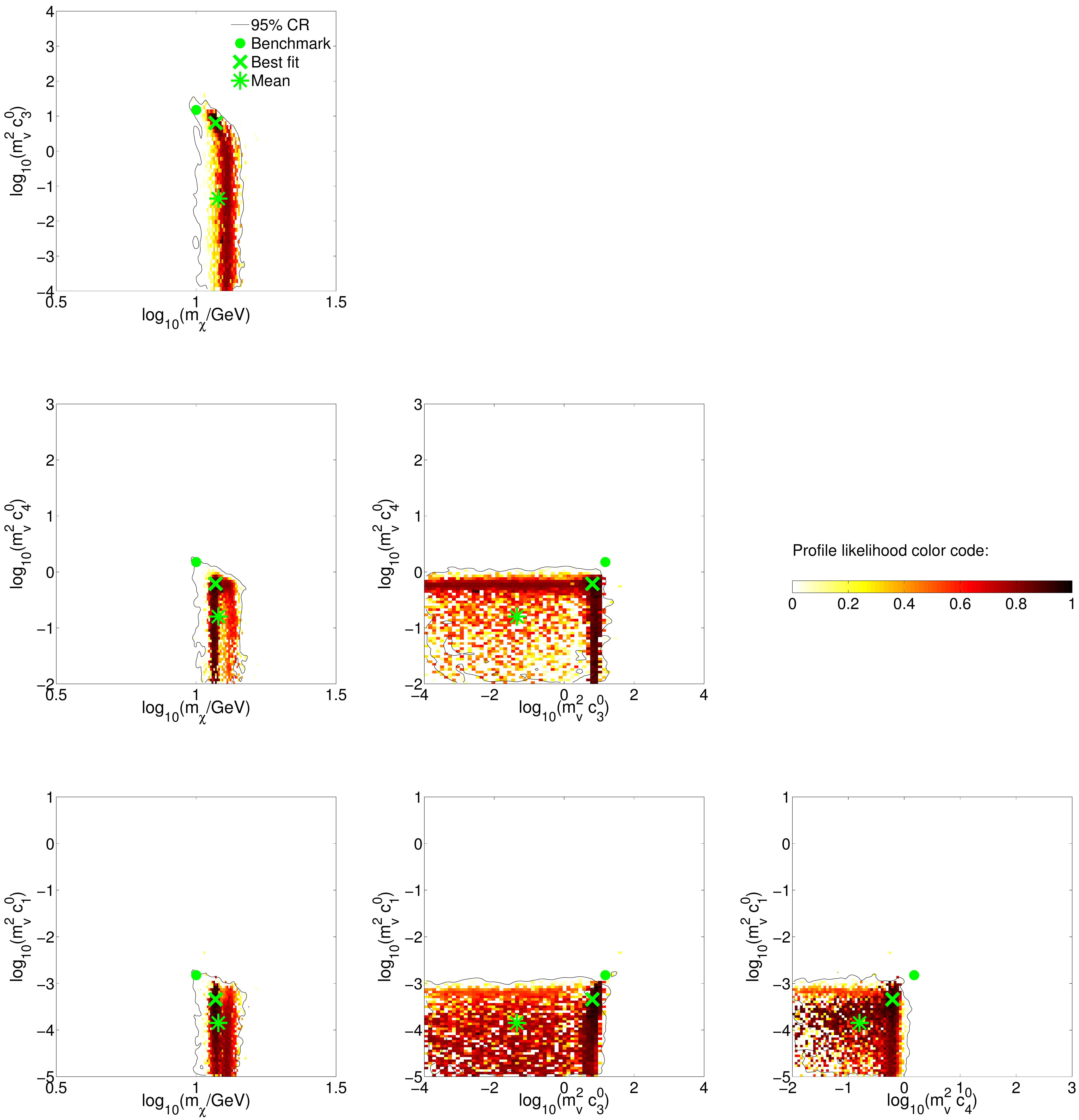}
\end{center}
\caption{Same as Fig.~\ref{fig:P26}, but for the benchmark points $P_{25}$.}
\label{fig:P25}
\end{figure}
\begin{figure}[t]
\begin{center}
\includegraphics[width=\textwidth]{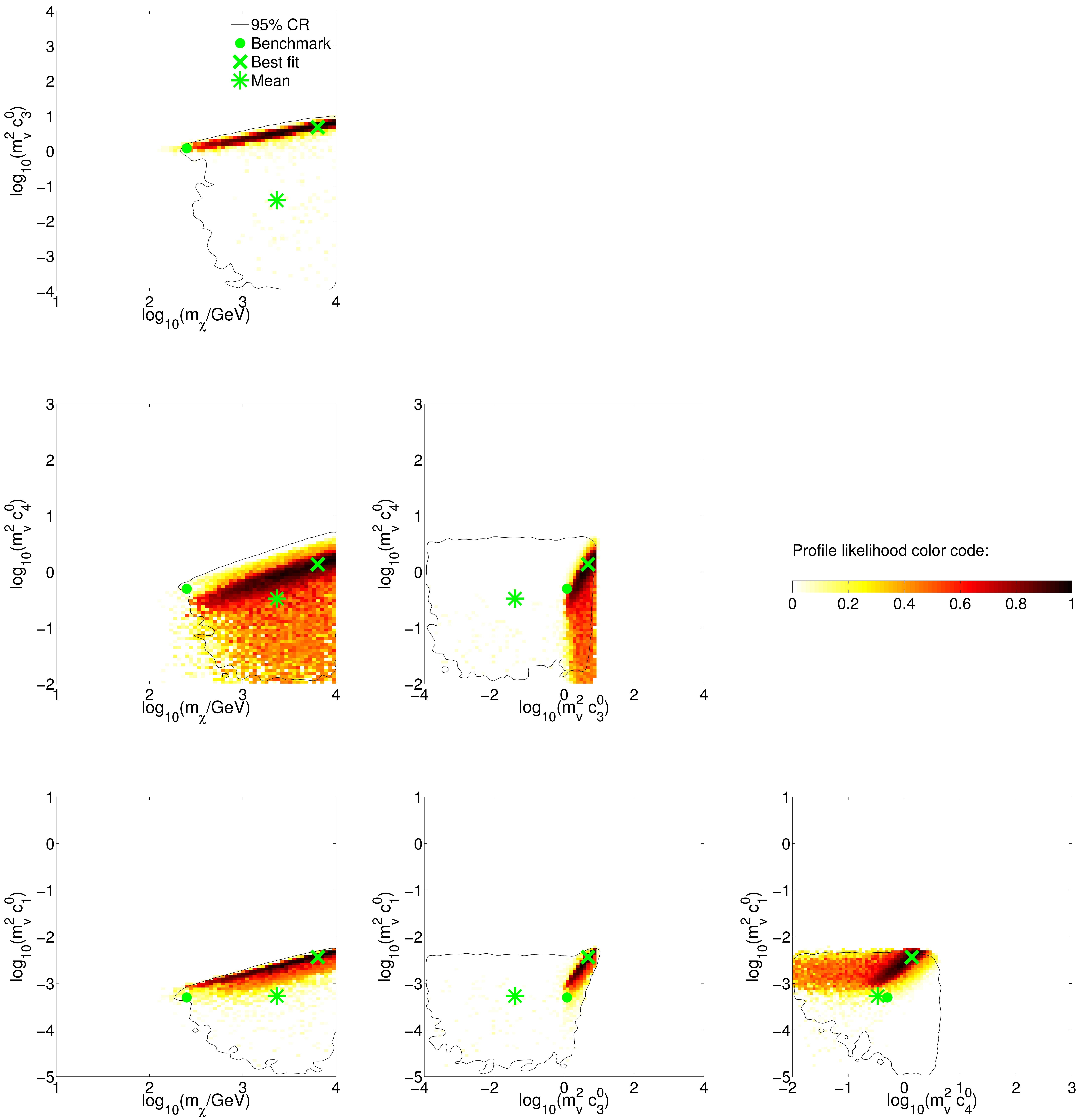}
\end{center}
\caption{Same as Fig.~\ref{fig:P26}, but for the benchmark points $P_{27}$.}
\label{fig:P27}
\end{figure}

Applying the fitting procedure B to the benchmark points $P_1$ -- $P_8$, we find that the correct dark matter mass can be in all cases extracted from our synthetic data. Remarkably, for the benchmark points $P_1$ -- $P_8$, the relative difference between the best fit value of $\log_{10}(m_\chi/{\rm GeV})$, and the value of $\log_{10}(m_\chi/{\rm GeV})$ at the benchmark point is of the order of a few percent (see Tab.~\ref{tab:results}). Extracting the correct dark matter-nucleon interaction operator from our synthetic data is however difficult in the case of the benchmark points $P_1$ -- $P_8$. The 2D profile likelihoods reported in Fig.~\ref{fig:P1P8} are indeed very flat, and different interaction operators can fit the data. From an analysis of the benchmark points $P_{16}$ -- $P_{24}$ based on the  fitting procedure B, we find that in the case of the point $P_{22}$, we can extract the correct dark matter mass and the correct dark matter-nucleon interaction type from the synthetic data. The relative difference between $\log_{10}(m_\chi/{\rm GeV})$ at the best fit point, and $\log_{10}(m_\chi/{\rm GeV})$ at $P_{22}$ is of about 15\%. Analogously to the benchmark point $P_{14}$, at $P_{22}$ the operator $\mathcal{O}_9 = -i\vec{S}_\chi\cdot(\vec{S}_N\times\vec{q}/m_N)$ is the leading dark matter-nucleon interaction operator. The properties of the remaining benchmark points with $m_{\chi}=250$~GeV are more difficult to extract from the synthetic data randomly generated in this study. 

Figs.~\ref{fig:P9P16}, \ref{fig:P1P8} and \ref{fig:P17P24} also show the 2D marginal posterior PDFs extracted from our synthetic data, and the associated 95\% CR contours (black lines in the figures). The 95\% CR contours follow the support of the 2D profile likelihoods, except for the benchmark points $P_9$, $P_{10}$, $P_{12}$, $P_{15}$ and $P_{16}$. In general, the Bayesian means in Figs.~\ref{fig:P9P16}, \ref{fig:P1P8} and \ref{fig:P17P24} (green stars in the figures) deviate from the benchmark points considerably. Indeed, because of volume effects, Bayesian means tend to prefer values of $m_\chi$ corresponding to regions in parameter space where the posterior PDF is relatively small.

Tab.~\ref{tab:results} (at the end of the paper) provides a detailed summary of our statistical analyses based on the fitting procedure B. For each benchmark point, Tab.~\ref{tab:results} shows the best fit values of the model parameters different from zero at the benchmark point, and the 95\% CL upper limits on the remaining model parameters.

\subsection{Prospects for benchmark points $P_{25}$ -- $P_{27}$}
\label{sec:P25P27}

In the last part of this section, we analyze the synthetic data randomly generated from the benchmark points $P_{25}$ -- $P_{27}$. At the benchmark points $P_{25}$ -- $P_{27}$, dark matter interacts with the nucleons through a linear combination of operators $\mathcal{O}_1$,  $\mathcal{O}_3$ and $\mathcal{O}_4$. The operators $\mathcal{O}_1$ and $\mathcal{O}_4$ are the leading interaction operators in a velocity/momentum power series expansion. Interestingly, the operator $\mathcal{O}_3$ interferes with the operator $\mathcal{O}_1$, as one can see by inspection of the dark matter response functions listed in the Appendix. See also Ref.~\cite{Catena:2014uqa}, for a detailed analysis of the $\mathcal{O}_1$-$\mathcal{O}_3$ correlation. 

Fig.~\ref{fig:P26} shows the 2D profile likelihoods that we find applying the fitting procedure B to the synthetic data generated from the benchmark point $P_{26}$. Different panels refer to distinct pairs of model parameters. More specifically, we show the 2D profile likelihoods in the six planes $m_\chi$-$c_1^0$,  $m_\chi$-$c_3^0$, $m_\chi$-$c_4^0$,  $c_3^0$-$c_1^0$, $c_3^0$-$c_4^0$ and $c_4^0$-$c_1^0$. In Fig.~\ref{fig:P26} we also report the 95\% CR contours obtained from a Bayesian analysis of the same dataset (black lines). The green circle, cross and star in each panel represent the benchmark point, the best fit point and the Bayesian mean, respectively. 

Comparing the position of the benchmark point $P_{26}$ projected in the six panels of Fig.~\ref{fig:P26} with the one of the best fit point in each panel, we see that, in the case of $P_{26}$, the dark matter mass and coupling constants can be simultaneously reconstructed applying the fitting procedure B to our synthetic data. This conclusion is confirmed by an analysis of the best fit values listed in Tab.~\ref{tab:results}. For instance, at the best fit point we find $\log_{10}(m_\chi/{\rm GeV})=1.86$ and $\log_{10}(m_v^2 c_1^0)=-3.39$, whereas at the benchmark point $P_{26}$,  $\log_{10}(m_\chi/{\rm GeV})=1.70$ and $\log_{10}(m_v^2 c_1^0)=-3.60$. This is a remarkable result, since the fitting procedure B does not assume any prior information on the form of the dark matter-nucleon interaction. The reconstruction of the correlated coupling constants $c_1^0$ and $c_3^0$ is particularly effective, as shown by the 2D profile likelihood in the bottom-central panel of Fig.~\ref{fig:P26}. In addition, in each panel of Fig.~\ref{fig:P26} the Bayesian means do not differ from the best fit points dramatically. 

Figs.~\ref{fig:P25} and \ref{fig:P27} show our findings for the benchmark points $P_{25}$ and $P_{27}$, respectively. In the case of $P_{25}$ (i.e. a light dark matter candidate) the best fit $m_\chi$ is larger than the value of $m_\chi$ at the benchmark point, whereas the best fit coupling constants are systematically smaller than their values at $P_{25}$. The relative difference between $\log_{10}(m_\chi/{\rm GeV})$ at the best fit and at the benchmark point is of about 10\%. However, the value of $m_\chi$ at the benchmark point is not contained in the 1D interval at 95\% CL extracted from the profile Likelihood by the  {\sffamily Getplots} program (see Tab.~\ref{tab:results}). In the case of $P_{27}$ (i.e. a relatively heavy dark matter candidate) the benchmark point reconstruction procedure appears to be more complicated, given the synthetic data generated in this study. The best fit dark matter mass and coupling constants are significantly larger than at the benchmark point $P_{27}$.

\section{Conclusions}

We have studied the prospects for direct detection of dark matter with future ton-scale detectors in the 11-dimensional effective theory of isoscalar dark matter-nucleon interactions mediated by a heavy spin-one or spin-zero particle. 

From a variegated sample of 27 benchmark points selected in the parameter space of the theory, we have simulated independent sets of synthetic data for ton-scale Germanium and Xenon detectors. We have then analyzed the synthetic data, using state-of-the-art Bayesian and frequentist statistical methods and numerical tools. From our statistical analyses, we have extracted 2D marginal posterior PDFs and profile likelihoods in the planes spanned by the independent pairs of model parameters. We have also computed the best fit points (a frequentist approach) and the means (a Bayesian approach) of the model parameters. 

Comparing the best fit points and the means of the model parameters to their benchmark values, we have identified the most promising scenarios and the main challenges emerging from our analysis of the 27 benchmark points. For all benchmark points, we find that the correct dark matter mass and coupling constants can be extracted from the synthetic data, when in the fit the true dark matter-nucleon interaction type is assumed a priori. This fitting procedure corresponds to the usual way of fitting the familiar spin-independent and spin-dependent interactions to direct detection data. 

The problem of extracting the correct dark matter-nucleon interaction type from the data directly is of a much more complex nature. It requires the exploration of the full 11-dimensional parameter space of the dark matter-nucleon effective theory. In our analyses, we have identified five scenarios where the dark matter mass, the dark matter-nucleon interaction type and the associated coupling constant can be extracted from the synthetic data simultaneously. The five scenarios correspond to the benchmark points $P_{10}$, $P_{11}$, $P_{14}$, and $P_{26}$, with $m_\chi=50$~GeV, and to the benchmark point $P_{22}$, with $m_\chi=250$~GeV. At the benchmark points $P_{10}$ and $P_{11}$ the leading dark matter-nucleon interaction operators are $\mathcal{O}_5 = -i\vec{S}_\chi\cdot(\vec{q}/m_N\times\vec{v}^{\perp}_{\chi N})$ and $\mathcal{O}_6 = (\vec{S}_\chi\cdot\vec{q}/m_N)(\vec{S}_N\cdot\vec{q}/m_N)$, respectively. The benchmark points $P_{14}$ and $P_{22}$ assume $\mathcal{O}_9 = -i\vec{S}_\chi\cdot(\vec{S}_N\times\vec{q}/m_N)$ as leading interaction operator, whereas at the benchmark point $P_{26}$, dark matter interacts with the nucleons through a linear combination of the interaction operators  $\mathcal{O}_1 = 1_{\chi} 1_{N}$,  $\mathcal{O}_3 = -i\vec{S}_N\cdot(\vec{q}/m_N\times\vec{v}^{\perp}_{\chi N})$ and $\mathcal{O}_4 = \vec{S}_{\chi}\cdot \vec{S}_{N}$. In the analyses of the benchmark points with $m_\chi=10$~GeV, we have always been able to accurately reconstruct the dark matter mass from the synthetic data, though it has been difficult to extract the correct dark matter-nucleon interaction type from the same data.  

We leave for future work the analysis of other detector types, and a study of the prospects for detecting dark matter-nucleon isovector interactions. Interestingly, isovector couplings probe nuclear response functions unexplored by the present analysis, whereas a larger suite of detectors would guarantee a better sensitivity to a dark matter induced recoil energy spectrum. 

In summary, we have presented the first comprehensive analysis of the prospects for direct detection of dark matter with future ton-scale detectors in the 11-dimensional effective theory of the dark matter-nucleon interaction. In our analyses, we have stressed the importance of extracting the correct dark matter nucleon-interaction type from the data directly, describing the challenges found addressing this complex 11-dimensional problem. The results presented in this paper will be particularly useful in interpreting the data of future dark matter direct detection experiments.

\appendix
\section{Dark matter response functions}
\label{sec:dmrfun}
In what follows we list the dark matter response functions appearing in Eq.~(\ref{Ptot}). These response functions have been derived from the ones of Ref.~\cite{Anand:2013yka}, setting to zero the couplings $c^{\tau}_{12},\dots,c^{\tau}_{15}$, with $\tau=0,1$:

\allowdisplaybreaks
\begin{eqnarray}
\label{eq:Rfunctions}
 R_{M}^{\tau \tau^\prime}({v}^{\perp 2}_{\chi T}, {{q}^{2} \over m_N^2}) &=& c_1^\tau c_1^{\tau^\prime } + {j_\chi (j_\chi+1) \over 3} \left[ {{q}^{2} \over m_N^2} {v}^{\perp 2}_{\chi T} c_5^\tau c_5^{\tau^\prime }+{v}^{\perp 2}_{\chi T} c_8^\tau c_8^{\tau^\prime }
+ {{q}^{2} \over m_N^2} c_{11}^\tau c_{11}^{\tau^\prime } \right] \nonumber \\
 R_{\Phi^{\prime \prime}}^{\tau \tau^\prime}({v}^{\perp 2}_{\chi T}, {{q}^{2} \over m_N^2}) &=& {{q}^{2} \over 4 m_N^2} c_3^\tau c_3^{\tau^\prime } 
 \nonumber \\
 R_{\Phi^{\prime \prime} M}^{\tau \tau^\prime}({v}^{\perp 2}_{\chi T}, {{q}^{2} \over m_N^2}) &=&  c_3^\tau c_1^{\tau^\prime } 
 \nonumber \\
  R_{\tilde{\Phi}^\prime}^{\tau \tau^\prime}({v}^{\perp 2}_{\chi T}, {{q}^{2} \over m_N^2}) &=& 0
  \nonumber \\
   R_{\Sigma^{\prime \prime}}^{\tau \tau^\prime}({v}^{\perp 2}_{\chi T}, {{q}^{2} \over m_N^2})  &=&{{q}^{2} \over 4 m_N^2} c_{10}^\tau  c_{10}^{\tau^\prime } +
  {j_\chi (j_\chi+1) \over 12} \left[ c_4^\tau c_4^{\tau^\prime} + 
 {{q}^{2} \over m_N^2} ( c_4^\tau c_6^{\tau^\prime }+c_6^\tau c_4^{\tau^\prime })+
 {{q}^{4} \over m_N^4} c_{6}^\tau c_{6}^{\tau^\prime } \right] \nonumber \\
    R_{\Sigma^\prime}^{\tau \tau^\prime}({v}^{\perp 2}_{\chi T}, {{q}^{2} \over m_N^2})  &=&{1 \over 8} \left[ {{q}^{2} \over  m_N^2}  {v}^{\perp 2}_{\chi T} c_{3}^\tau  c_{3}^{\tau^\prime } + {v}^{\perp 2}_{\chi T}  c_{7}^\tau  c_{7}^{\tau^\prime }  \right]
       + {j_\chi (j_\chi+1) \over 12} \left[ c_4^\tau c_4^{\tau^\prime} + 
        {{q}^{2} \over m_N^2} c_9^\tau c_9^{\tau^\prime }
        \right] \nonumber \\
     R_{\Delta}^{\tau \tau^\prime}({v}^{\perp 2}_{\chi T}, {{q}^{2} \over m_N^2})&=&  {j_\chi (j_\chi+1) \over 3} \left[ {{q}^{\,2} \over m_N^2} c_{5}^\tau c_{5}^{\tau^\prime }+ c_{8}^\tau c_{8}^{\tau^\prime } \right] \nonumber \\
 R_{\Delta \Sigma^\prime}^{\tau \tau^\prime}({v}^{\perp 2}_{\chi T}, {{q}^{2} \over m_N^2})&=& {j_\chi (j_\chi+1) \over 3} \left[c_{5}^\tau c_{4}^{\tau^\prime }-c_8^\tau c_9^{\tau^\prime} \right].
\end{eqnarray}
We assume for definiteness that the dark matter particle has spin $j_\chi=1/2$.

\begin{table}[t]
\tiny
    \centering
    \begin{tabular}{lccccccccccc}
    \toprule
    &$P_{1}(c_3^0)$ & $P_{2}(c_5^0)$ &  $P_{3}(c_6^0)$ &  $P_{4}(c_7^0)$ &  $P_{5}(c_8^0)$ &  $P_{6}(c_9^0)$ &  $P_{7}(c_{10}^0)$ &  $P_{8}(c_{11}^0)$  \\
    \midrule  
    $\log_{10}(m_\chi/{\rm GeV})$~[BP]  & 1 &  1 & 1 & 1 & 1 & 1 & 1 & 1 \\        
    $\log_{10}(m_\chi/{\rm GeV})$~[BF]  & 1.02 & 1.03 & 1.03 &  0.98 & 0.97 & 1.03 & 1.04 & 1.03 \\     
    $\log_{10}(m_\chi/{\rm GeV})$~[CL]   &[0.98,1.05] & [0.99,1.06] & [1.00,1.05] & [0.95,1.02] & [0.96,1.02] & [0.99,1.04] & [1.00,1.06]  & [0.99,1.05]  \\ 
    $\log_{10}(G_i\neq 0)$~[BP]   & 1.30 &  2.08 & 3.48 &3.41&0.78 & 1.90&1.70 & -1.00  \\ 
    $\log_{10}(G_i\neq 0)$~[BF]  & -2.41 & 1.67 & -2.26 & 2.86 & -0.33 & 0.81 & -0.24 & -3.08\\ 
    $\log_{10}(G_i\neq 0)$~[CL] &  [-4,1.49] & [-4,2.16] & [-4,3.32] & [-4,3.35]& [-4, 0.44]& [-4,2.03] &[-4,1.63] &[-4,-0.94]  \\                                  
    $\log_{10}(G_1)~[{\rm UL}]$ & -2.43 & -2.59 & -2.10 & -2.73 & -2.74 & -2.65 &-2.47 & -2.44  \\   
    $\log_{10}(G_3)~[{\rm UL}]$  & 1.49 & 1.29 & 1.82 & 1.49 & 1.31 & 1.11 &  1.45 &  1.57 \\ 
    $\log_{10}(G_4)~[{\rm UL}]$  & 0.36 & 0.26 & 0.30 & 0.26 & 0.17 & 0.35 & 0.15 & 0.23 \\ 
    $\log_{10}(G_5)~[{\rm UL}]$  & 2.34 & 2.16 & 2.18 & 2.42 & 2.39 & 2.26 & 2.03 & 2.11   \\ 
    $\log_{10}(G_6)~[{\rm UL}]$  & 3.60 & 3.47 & 3.32 & 3.60 & 3.68 & 3.47 & 3.42 & 3.51 \\ 
    $\log_{10}(G_7)~[{\rm UL}]$  & 3.36 & 3.34 & 3.39 & 3.35 & 3.01 &3.42 & 3.13 & 3.16  \\ 
    $\log_{10}(G_8)~[{\rm UL}]$  & 0.76 & 0.59 & 0.66  & 0.58 & 0.44 &0.68 &0.48 & 0.54 \\ 
    $\log_{10}(G_9)~[{\rm UL}]$  & 1.99 & 1.91 & 1.90 & 2.14 & 2.03 & 2.03 & 1.90 &1.83  \\ 
    $\log_{10}(G_{10})~[{\rm UL}]$  & 1.78 & 1.74 &1.63 & 1.95 & 1.91 & 1.81  &1.63 & 1.68 \\ 
    $\log_{10}(G_{11})~[{\rm UL}]$  & -0.78 & -0.93 & -0.94 & -0.69 & -0.74 &-0.87 & -0.99 &-0.94 \\ 
     \midrule  
      &$P_{9}(c_3^0)$ & $P_{10}(c_5^0)$ &  $P_{11}(c_6^0)$ &  $P_{12}(c_7^0)$ &  $P_{13}(c_8^0)$ &  $P_{14}(c_{9}^0)$ &  $P_{15}(c_{10}^0)$ &  $P_{16}(c_{11}^0)$  \\
     \midrule
     $\log_{10}(m_\chi/{\rm GeV})$~[BP] & 1.7 & 1.7& 1.7& 1.7&1.7&1.7 & 1.7& 1.7 \\        
    $\log_{10}(m_\chi/{\rm GeV})$~[BF]  & 1.75 & 1.67&1.86& 1.44 & 2.00& 1.71& 1.72& 1.66\\     
    $\log_{10}(m_\chi/{\rm GeV})$~[CL] &[1.69,2.04] &[1.59,1.85]& [1.75,2.41]& [1.36,1.52]& [1.60,3]& [1.60,2.98]& [1.63,1.99]&  [1.61,1.95]\\ 
    $\log_{10}(G_i\neq 0)$~[BP]  & -0.097 & 1.08 & 2.0 &2.48&  -0.30 & 0.90 & 0.60 &-2.0   \\ 
    $\log_{10}(G_i\neq 0)$~[BF]   & -1.70& 1.09 & 1.81&-0.77& -2.14& 0.89 &-3.83& -2.65\\ 
    $\log_{10}(G_i\neq 0)$~[CL]   &[-4,-0.15] & [-4,1.25]& [-4,2.08]& [-4,2.20]& [-4,-0.18]& [-4,1.05]& [-4,0.48]&  [-4,-1.91]\\                                  
    $\log_{10}(G_1)~[{\rm UL}]$   &-3.81& -3.48& -3.83 &-3.67& -2.84 & -2.84 &-3.55& -3.53 \\  
    $\log_{10}(G_3)~[{\rm UL}]$ &-0.15& -0.071 & -0.13 &-0.23& 0.09 & -0.04 &-0.22&  -0.17\\  
    $\log_{10}(G_4)~[{\rm UL}]$  &-0.83&-0.43& -0.83 & -0.81 &-0.25 & -0.26 & -0.57 &-0.49 \\ 
    $\log_{10}(G_5)~[{\rm UL}]$   &1.12 & 1.25 &1.02 & 0.87 & 0.93& 1.06 & 1.05 & 1.19  \\ 
    $\log_{10}(G_6)~[{\rm UL}]$  &2.01& 2.09 &1.92 &1.80 & 1.70 & 1.82 & 1.83 & 2.03 \\ 
    $\log_{10}(G_7)~[{\rm UL}]$   &2.10&  2.50 &2.08 &2.20& 2.88 & 2.88 & 2.41 & 2.35 \\ 
    $\log_{10}(G_8)~[{\rm UL}]$  &-0.55&  -0.18 &-0.43 &-0.18  &-0.18 &-0.11& -0.24 & -0.26 \\ 
    $\log_{10}(G_9)~[{\rm UL}]$   &0.87&  0.99 & 0.96 & 1.02 & 0.94 &1.05& 1.05& 1.07\\ 
    $\log_{10}(G_{10})~[{\rm UL}]$   &0.65& 0.69 & 0.61 & 0.43 &0.38& 0.46&  0.48& 0.55\\ 
    $\log_{10}(G_{11})~[{\rm UL}]$ &-1.90& -1.84 & -1.91 & -2.13 & -2.15 &-2.03&-1.98& -1.91\\      \midrule  
     &$P_{17}(c_3^0)$ & $P_{18}(c_5^0)$ &  $P_{19}(c_6^0)$ &  $P_{20}(c_7^0)$ &  $P_{21}(c_8^0)$ &  $P_{22}(c_{9}^0)$ &  $P_{23}(c_{10}^0)$ &  $P_{24}(c_{11}^0)$  \\
     \midrule
   $\log_{10}(m_\chi/{\rm GeV})$~[BP]  & 2.4&2.4 &2.4 &2.4& 2.4& 2.4&2.4 & 2.4  \\        
    $\log_{10}(m_\chi/{\rm GeV})$~[BF] &3.87& 3.63& 4& 1.74 &1.75 &2.11& 2.26 & 2.13 \\     
    $\log_{10}(m_\chi/{\rm GeV})$~[CL]   & [2.20,4] & [2.25,4]& [2.31,4]& [1.57,4]& [1.55,4]& [1.82,4]& [1.87,4] & [1.88,4]\\ 
    $\log_{10}(G_i\neq 0)$~[BP]   & -0.097 & 1.08 & 2.0 &2.48&  -0.30 & 0.90 & 0.60 &-2.0   \\ 
    $\log_{10}(G_i\neq 0)$~[BF]   & 0.71& 1.49&-1.92& -0.36 & -1.42 &0.79& -1.58 & -2.40\\ 
    $\log_{10}(G_i\neq 0)$~[CL]   & [-4,0.86]& [-4,1.85]& [-4,2.82]& [-4,3.47]& [-4,0.34]& [-4,1.79]& [-4,1.39] & [-4,-1.23] \\                                  
    $\log_{10}(G_1)~[{\rm UL}]$  & -2.68 &-2.63& -2.62& -2.62 & -2.56 &-2.58& -2.61 & -2.68\\      
    $\log_{10}(G_3)~[{\rm UL}]$  & 0.86 & 0.49& 0.89& 0.17 & 0.28 &  0.24 & 0.42 & 0.18 \\ 
    $\log_{10}(G_4)~[{\rm UL}]$   & 0.35 & 0.20 & 0.29 & 0.38  &0.35 &  0.35 & 0.39 & 0.32\\ 
    $\log_{10}(G_5)~[{\rm UL}]$  & 1.80 &1.85& 1.80 & 1.40 & 1.24 & 1.42 & 1.72 & 1.35\\ 
    $\log_{10}(G_6)~[{\rm UL}]$  &2.84 &2.66& 2.82 & 1.67 & 2.15 & 2.10 & 2.47 & 2.20\\ 
    $\log_{10}(G_7)~[{\rm UL}]$  &3.21 &3.15& 3.15 & 3.47 & 3.43 & 3.32  & 3.42 & 3.27\\ 
    $\log_{10}(G_8)~[{\rm UL}]$  &0.58 & 0.72& 0.55 & 0.19 & 0.34 & 0.71 & 0.76 & 0.66\\ 
    $\log_{10}(G_9)~[{\rm UL}]$ &1.65 & 1.96& 1.71 &1.15 &1.52  & 1.79 & 1.88 & 1.88\\ 
    $\log_{10}(G_{10})~[{\rm UL}]$   & 1.48& 1.36&  1.43 &0.75 &1.00& 1.07 & 1.39 & 1.05 \\ 
    $\log_{10}(G_{11})~[{\rm UL}]$  & -1.27 & -1.07& -1.24 &-1.81 &-1.68 &  -1.33 & -1.13  & -1.23 \\ 
    \midrule
     & & $P_{25}$  &  & $P_{26}$  &  &   $P_{27}$ & &   \\
     \midrule
   $\log_{10}(m_\chi/{\rm GeV})$~[BP]  & & 1 && 1.7 & & 2.4&  \\        
    $\log_{10}(m_\chi/{\rm GeV})$~[BF] & & 1.07 & & 1.86 & & 3.80  &  \\     
    $\log_{10}(m_\chi/{\rm GeV})$~[CL]   &  & [1.02,1.18]& & [1.58,3]& & [2.23,4]  & \\ 
    $\log_{10}(G_1)$~[BP]  & &  -2.82 && -3.60 & & -3.30 &  \\ 
    $\log_{10}(G_1)$~[BF]  & & -3.34 && -3.39 & & -2.43 &  \\  
    $\log_{10}(G_3)$~[BP]  & & 1.18 &&  -0.15 & & 0.079&  \\ 	
    $\log_{10}(G_3)$~[BF]  & & 0.81 &&  -0.18 & & 0.68 &  \\  
    $\log_{10}(G_4)$~[BP]  & &0.18 &&  -0.60 & & -0.30 &  \\ 
    $\log_{10}(G_4)$~[BF]  & & -0.21 && -0.73 & & 0.13 &  \\ 
    $\log_{10}(G_1)~[{\rm UL}]$  & & -2.34 && -2.83 & & -2.27 &  \\ 
    $\log_{10}(G_3)~[{\rm UL}]$  & & 1.59 && 0.26 & & 0.87 &  \\ 
    $\log_{10}(G_4)~[{\rm UL}]$   & & -0.03 && -0.27 & & 0.57 &  \\ 
    $\log_{10}(G_5)~[{\rm UL}]$  & & 1.83 && 0.83 & & 1.88&  \\ 
    $\log_{10}(G_6)~[{\rm UL}]$  & & 3.13 && 1.74 & & 2.80&  \\ 
    $\log_{10}(G_7)~[{\rm UL}]$  & & 2.98 && 2.82 & & 3.61 &  \\ 
    $\log_{10}(G_8)~[{\rm UL}]$  & & 0.31 && -0.23 & & 0.83 &  \\ 
    $\log_{10}(G_9)~[{\rm UL}]$ & & 1.53 && 0.76 & & 1.66 &  \\ 
    $\log_{10}(G_{10})~[{\rm UL}]$   & &1.45  && 0.42 & & 1.38 &  \\ 
    $\log_{10}(G_{11})~[{\rm UL}]$  & & -1.20 && -2.09 & & -1.09 &  \\ 
     \bottomrule
    \end{tabular}
    \normalsize
    \caption{Detailed summary of our statistical analyses based on the fitting procedure B. For each benchmark point (BP), this table shows the best fit (BF) values of the model parameters different from zero at the benchmark point, and the 95\% confidence level upper limits (UP) on the remaining model parameters. For certain parameters, we also report the 1D intervals at 95\% confidence level (CL) constructed from the profile Likelihood with {\sffamily Getplots}. To simplify the notation we have introduced the dimensionless quantities $G_i\equiv m_v^2 c_i^0$. }
    \label{tab:results}
\end{table}

\acknowledgments We would like to thank Paolo Gondolo for many inspiring conversations on the effective theory of the dark matter-nucleon interaction. R.C. acknowledges partial support from the European Union FP7 ITN INVISIBLES (Marie Curie Actions, PITN-GA-2011-289442).

\providecommand{\newblock}{}


\begin{thebibliography}{10}
\expandafter\ifx\csname url\endcsname\relax
  \def\url#1{{\tt #1}}\fi
\expandafter\ifx\csname urlprefix\endcsname\relax\def\urlprefix{URL }\fi
\providecommand{\eprint}[2][]{\url{#2}}

\bibitem{appec}
APPEC \urlprefix\url{http://www.appec.org/}

\bibitem{Strigari:2013iaa}
Strigari L~E 2013 {\em Phys.Rept.\/} {\bf 531} 1--88 (\textit{Preprint}
  \eprint{1211.7090})

\bibitem{Goodman:1984dc}
Goodman M~W and Witten E 1985 {\em Phys.Rev.\/} {\bf D31} 3059

\bibitem{Cerdeno:2014uga}
Cerdeno D, Cuesta C, Fornasa M, Garcia E, Ginestra C {\em et~al.\/} 2014
  (\textit{Preprint} \eprint{1403.3539})

\bibitem{Catena:2014uqa}
Catena R and Gondolo P 2014  (\textit{Preprint} \eprint{1405.2637})

\bibitem{Akerib:2013tjd}
Akerib D {\em et~al.\/} (LUX Collaboration) 2013  (\textit{Preprint}
  \eprint{1310.8214})

\bibitem{Agnese:2014aze}
Agnese R {\em et~al.\/} (SuperCDMS Collaboration) 2014  (\textit{Preprint}
  \eprint{1402.7137})

\bibitem{Agnese:2013jaa}
Agnese R {\em et~al.\/} (SuperCDMSSoudan Collaboration) 2014 {\em
  Phys.Rev.Lett.\/} {\bf 112} 041302 (\textit{Preprint} \eprint{1309.3259})

\bibitem{Baudis:2012ig}
Baudis L 2012 {\em Phys.Dark Univ.\/} {\bf 1} 94--108 (\textit{Preprint}
  \eprint{1211.7222})

\bibitem{Akrami:2010dn}
Akrami Y, Savage C, Scott P, Conrad J and Edsjo J 2011 {\em JCAP\/} {\bf 1104}
  012 (\textit{Preprint} \eprint{1011.4318})

\bibitem{Pato:2011de}
Pato M 2011 {\em JCAP\/} {\bf 1110} 035 (\textit{Preprint} \eprint{1106.0743})

\bibitem{Strege:2012kv}
Strege C, Trotta R, Bertone G, Peter A~H and Scott P 2012 {\em Phys.Rev.\/}
  {\bf D86} 023507 (\textit{Preprint} \eprint{1201.3631})

\bibitem{Catena:2013pka}
Catena R and Covi L 2013  (\textit{Preprint} \eprint{1310.4776})

\bibitem{Green:2008rd}
Green A~M 2008 {\em JCAP\/} {\bf 0807} 005 (\textit{Preprint}
  \eprint{0805.1704})

\bibitem{Trotta:2006ew}
Trotta R, de~Austri R~R and Roszkowski L 2007 {\em New Astron. Rev.\/} {\bf 51}
  316--320 (\textit{Preprint} \eprint{astro-ph/0609126})

\bibitem{Akrami:2010cz}
Akrami Y, Savage C, Scott P, Conrad J and Edsjo J 2011 {\em JCAP\/} {\bf 1107}
  002 (\textit{Preprint} \eprint{1011.4297})

\bibitem{Bergstrom:2010gh}
Bergstrom L, Bringmann T and Edsjo J 2011 {\em Phys.Rev.\/} {\bf D83} 045024
  (\textit{Preprint} \eprint{1011.4514})

\bibitem{Pato:2010zk}
Pato M, Baudis L, Bertone G, Ruiz~de Austri R, Strigari L~E {\em et~al.\/} 2011
  {\em Phys.Rev.\/} {\bf D83} 083505 (\textit{Preprint} \eprint{1012.3458})

\bibitem{Arina:2013jma}
Arina C 2013  (\textit{Preprint} \eprint{1310.5718})

\bibitem{Peter:2013aha}
Peter A~H~G, Gluscevic V, Green A~M, Kavanagh B~J and Lee S~K 2013
  (\textit{Preprint} \eprint{1310.7039})

\bibitem{Gondolo:2002np}
Gondolo P 2002 {\em Phys.Rev.\/} {\bf D66} 103513 (\textit{Preprint}
  \eprint{hep-ph/0209110})

\bibitem{Drees:2007hr}
Drees M and Shan C~L 2007 {\em JCAP\/} {\bf 0706} 011 (\textit{Preprint}
  \eprint{astro-ph/0703651})

\bibitem{Peter:2011eu}
Peter A~H 2011 {\em Phys.Rev.\/} {\bf D83} 125029 (\textit{Preprint}
  \eprint{1103.5145})

\bibitem{Bozorgnia:2014dqa}
Bozorgnia N and Schwetz T 2014  (\textit{Preprint} \eprint{1405.2340})

\bibitem{Fitzpatrick:2012ix}
Fitzpatrick A~L, Haxton W, Katz E, Lubbers N and Xu Y 2013 {\em JCAP\/} {\bf
  1302} 004 (\textit{Preprint} \eprint{1203.3542})

\bibitem{Gresham:2014vja}
Gresham M~I and Zurek K~M 2014  (\textit{Preprint} \eprint{1401.3739})

\bibitem{Fan:2010gt}
Fan J, Reece M and Wang L~T 2010 {\em JCAP\/} {\bf 1011} 042 (\textit{Preprint}
  \eprint{1008.1591})

\bibitem{Fitzpatrick:2012ib}
Fitzpatrick A~L, Haxton W, Katz E, Lubbers N and Xu Y 2012  (\textit{Preprint}
  \eprint{1211.2818})

\bibitem{Liang:2013dsa}
Liang Z~L and Wu Y~L 2014 {\em Phys.Rev.\/} {\bf D89} 013010 (\textit{Preprint}
  \eprint{1308.5897})

\bibitem{Panci:2014gga}
 Panci~P {\em Adv.~High Energy Phys.}  {\bf 2014} (2014) 681312
  (\textit{Preprint} \eprint{1402.1507})

\bibitem{Anand:2013yka}
Anand N, Fitzpatrick A~L and Haxton W 2013  (\textit{Preprint}
  \eprint{1308.6288})

\bibitem{DelNobile:2013sia}
Cirelli M, Del~Nobile E and Panci P 2013 {\em JCAP\/} {\bf 1310} 019
  (\textit{Preprint} \eprint{1307.5955})

\bibitem{Bozorgnia:2013pua}
Bozorgnia N, Catena R and Schwetz T 2013 {\em JCAP\/} {\bf 1312} 050
  (\textit{Preprint} \eprint{1310.0468})

\bibitem{Catena:2009mf}
Catena R and Ullio P 2010 {\em JCAP\/} {\bf 1008} 004 (\textit{Preprint}
  \eprint{0907.0018})

\bibitem{Catena:2011kv}
Catena R and Ullio P 2012 {\em JCAP\/} {\bf 1205} 005 (\textit{Preprint}
  \eprint{1111.3556})

\bibitem{Feng:2011vu}
Feng J~L, Kumar J, Marfatia D and Sanford D 2011 {\em Phys.Lett.\/} {\bf B703}
  124--127 (\textit{Preprint} \eprint{1102.4331})

\bibitem{2011JHEP...06..042F}
{Feroz} F, {Cranmer} K, {Hobson} M, {Ruiz de Austri} R and {Trotta} R 2011 {\em
  Journal of High Energy Physics\/} {\bf 6} 42 (\textit{Preprint}
  \eprint{1101.3296})

\bibitem{Feroz:2008xx}
Feroz F, Hobson M and Bridges M 2009 {\em Mon.Not.Roy.Astron.Soc.\/} {\bf 398}
  1601--1614 (\textit{Preprint} \eprint{0809.3437})

\bibitem{Feroz:2007kg}
Feroz F and Hobson M 2008 {\em Mon.Not.Roy.Astron.Soc.\/} {\bf 384} 449
  (\textit{Preprint} \eprint{0704.3704})

\bibitem{Feroz:2013hea}
Feroz F, Hobson M, Cameron E and Pettitt A 2013  (\textit{Preprint}
  \eprint{1306.2144})

\bibitem{Lewis:2002ah}
Lewis A and Bridle S 2002 {\em Phys. Rev.\/} {\bf D66} 103511
  (\textit{Preprint} \eprint{astro-ph/0205436})

\bibitem{Austri:2006pe}
de~Austri R~R, Trotta R and Roszkowski L 2006 {\em JHEP\/} {\bf 0605} 002
  (\textit{Preprint} \eprint{hep-ph/0602028})

\end{thebibliography}
\end{document}